\documentclass{aa}  

\usepackage{graphicx}
\usepackage{txfonts}
\begin{document}

\title{The GTC exoplanet transit spectroscopy survey X.}

\subtitle{Stellar spots versus Rayleigh scattering: the case of HAT-P-11b}

\author{F. Murgas\inst{1}\fnmsep\inst{2}
          \and
          G. Chen \inst{1}\fnmsep\inst{2}\fnmsep\inst{3}
          \and
          E. Pall\'{e}\inst{1}\fnmsep\inst{2}
          \and
          L. Nortmann\inst{1}\fnmsep\inst{2}
          \and
          G. Nowak\inst{1}\fnmsep\inst{2}
}

\institute{Instituto de Astrof\'isica de Canarias (IAC), E-38205 La Laguna, Tenerife, Spain\\
              \email{fmurgas@iac.es} 
              \and
              Departamento de Astrof\'isica, Universidad de La Laguna (ULL), E-38206 La Laguna, Tenerife, Spain
              \and
              Key Laboratory of Planetary Sciences, Purple Mountain Observatory, Chinese Academy of Sciences, Nanjing 210008, China
}

\date{Received August 10, 2018; accepted December 22, 2018}

 
  \abstract
  {Rayleigh scattering in a hydrogen-dominated exoplanet atmosphere can be detected from ground or space based telescopes, however, stellar activity in the form of spots can mimic Rayleigh scattering in the observed transmission spectrum. Quantifying this phenomena is key to our correct interpretation of exoplanet atmospheric properties.}
   {We use the 10 m telescope Gran Telescopio Canarias (GTC) to carry out a ground-based transmission spectra survey of extrasolar planets to characterize their atmospheres. In this paper we investigate the exoplanet HAT-P-11b, a Neptune-size planet orbiting an active K-type star.}
   {We obtained long-slit optical spectroscopy of two transits of HAT-P-11b with the Optical System for Imaging and low-Intermediate-Resolution Integrated Spectroscopy (OSIRIS) on August 30 2016 and September 25 2017. We integrated the spectrum of HAT-P-11 and one reference star in several spectroscopic channels across the $\lambda\sim$ 400-785 nm region, creating numerous light curves of the transits. We fit analytic transit curves to the data taking into account the systematic effects and red noise present in the time series in an effort to measure the change of the planet-to-star radius ratio ($R_\mathrm{p}/R_\mathrm{s}$) across wavelength.}
   {By fitting both transits together, we find a slope in the transmission spectrum showing an increase of the planetary radius towards blue wavelengths. A closer inspection to the transmission spectrum of the individual data sets reveals that the first transit presents this slope while the transmission spectrum of the second data set is flat. Additionally we detect hints of Na absorption in the first night, but not in the second. We conclude that the transmission spectrum slope and Na absorption excess found in the first transit observation are caused by unocculted stellar spots. Modeling the contribution of unocculted spots to reproduce the results of the first night we find a spot filling factor of $\delta=0.62^{+0.20}_{-0.17}$ and a spot-to-photosphere temperature difference of $\Delta T = 429^{+184}_{-299}$ K.}
   {}


   \maketitle
%

\section{Introduction}

A popular technique to study the atmospheres of transiting exoplanets is transmission spectroscopy. This method consists in measuring the change in the planetary radius across wavelengths during a transit event. In the past decade this procedure has been used to detect several planetary atmospheric features such as atomic and molecular absorption lines (e.g., \citealp{Charbonneau2002}, \citealp{Redfield2008}, \citealp{Snellen2008}, \citealp{Fraine2014}, \citealp{Kreidberg2014b}, \citealp{Chen2017b}, \citealp{Chen2018}, \citealp{Nikolov2018}, \citealp{Wakeford2018}) and scattering features (e.g., \citealp{LecavelierDesEtangs2008}, \citealp{Sing2011}, \citealp{Sing2013}, \citealp{Pont2013}, \citealp{Sing2015}, \citealp{Nikolov2015}, \citealp{Kirk2017}). Thanks to their relatively large atmospheric scale heights most of these detections have been accomplished for Jupiter analogs, however there has been several efforts in recent years to study the atmospheres of exoplanets with masses around the Saturn/Neptune regime and lower (e.g., \citealp{Bean2010}, \citealp{Stevenson2010}, \citealp{Fukui2013}, \citealp{Nascimbeni2013}, \citealp{Knutson2014}, \citealp{Kreidberg2014}, \citealp{Fraine2014}, \citealp{Dragomir2015}, \citealp{Tsiaras2016}, \citealp{Chen2017}, \citealp{DiamondLowe2018}). 

Despite the proven success of transmission spectroscopy, there is still a risk of attributing to the planet certain features whose origin may be on star. One of these features is Rayleigh scattering. Stellar activity in the form of spots can introduce signals that mimic the effect of an increase of the measured planetary radius toward the blue (e.g., \citealp{Pont2013}, \citealp{McCullough2014}, \citealp{Oshagh2014}).

The exoplanet HAT-P-11b is a Neptune-size planet discovered by \cite{Bakos2010} orbiting a relatively bright ($V = 9.47$ mag) active K dwarf star (\citealp{Morris2017} and references therein) with a period of 4.88 days. With an initial measured radius of $R_p = 0.422 \pm 0.014\; R_J$ ($R_p = 4.73 \pm 0.16\; R_\oplus$) and a mass of $M_p = 0.081 \pm 0.009\; M_J$ ($M_p = 25.8 \pm 2.9\; M_\oplus$), HAT-P-11b had the smallest radius discovered by a ground based survey at the time of discovery. Using \emph{Kepler} data, \cite{Deming2011} improved the planetary radius measurement to $R_p = 4.31 \pm 0.06\; R_\oplus$, a smaller radius than the reported in the discovery. \cite{Winn2010} took radial velocity (RV) measurements during a transit and found that the orbit of HAT-P-11b is highly misaligned with respect to the rotational axis of its host star with an angle of $103^{+26}_{-10}$ deg. Using Subaru RV data \cite{Hirano2011} found an angle $103^{+22}_{-18}$ deg, in agreement with \cite{Winn2010}. \cite{Lecavelier2013} studied the radio emission of the system at 150 MHz during an eclipse and detected radio emission which they attributed to the planet, although it has not been confirmed by follow up observations to date. \cite{Huber2017} detected the secondary eclipse of this planet using data from \emph{Kepler}, obtaining an eccentricity of $e=0.26459^{+0.00069}_{-0.00048}$ and a geometrical albedo similar to Neptune of $0.39\pm 0.07$. Using a decade of RV measurements, \cite{Yee2018} found a second planet orbiting HAT-P-11; with a mass similar to Jupiter ($M_p\sin i=1.6\pm 0.1; M_J$), a period of 9.3 years, and an eccentricity of $e=0.6\pm0.03$; HAT-P-11c could explain the orbital tilt of HAT-P-11b.

\cite{Fraine2014} obtained a transmission spectrum of HAT-P-11b using Hubble Space Telescope (HST) instrument Wide Field Camera 3 (WFC3) in the wavelength range of 1.1-1.7 $\mu$m coupled with \textit{Spitzer} observations at 3.6 and 4.5 $\mu$m. The data showed evidence of water absorption at 1.4 $\mu$m. \cite{Tsiaras2018} re-analyzed HST/WFC3 transit observations of several gaseous exoplanets, including the HAT-P-11b data set taken by \cite{Fraine2014} confirming the absorption feature of H$_2$O.

Here we present the optical transmission spectrum of HAT-P-11b obtained with the 10.4 m Gran Telescopio Canarias (GTC). This paper is organized as follows. In \S \ref{Sec:Obs} we describe the observations and data reduction, in \S \ref{Sec:DataRed} we describe the light curve fitting process, in \S \ref{Sec:Disc} we show the transmission spectrum of HAT-P-11b and discuss our results. Finally in \S \ref{Sec:Conclusions} we present the conclusions of this work.

\section{Observations and data reduction}
\label{Sec:Obs}
The data were taken with the 10.4 m telescope GTC at the Observatorio Roque de los Muchachos, in Spain. Using its Optical System for Imaging and low-Resolution-Integrated Spectroscopy (OSIRIS, \citealp{Cepa2000}) instrument in its long-slit spectroscopic mode.

\begin{figure}
  \centering
  \includegraphics[width=\hsize]{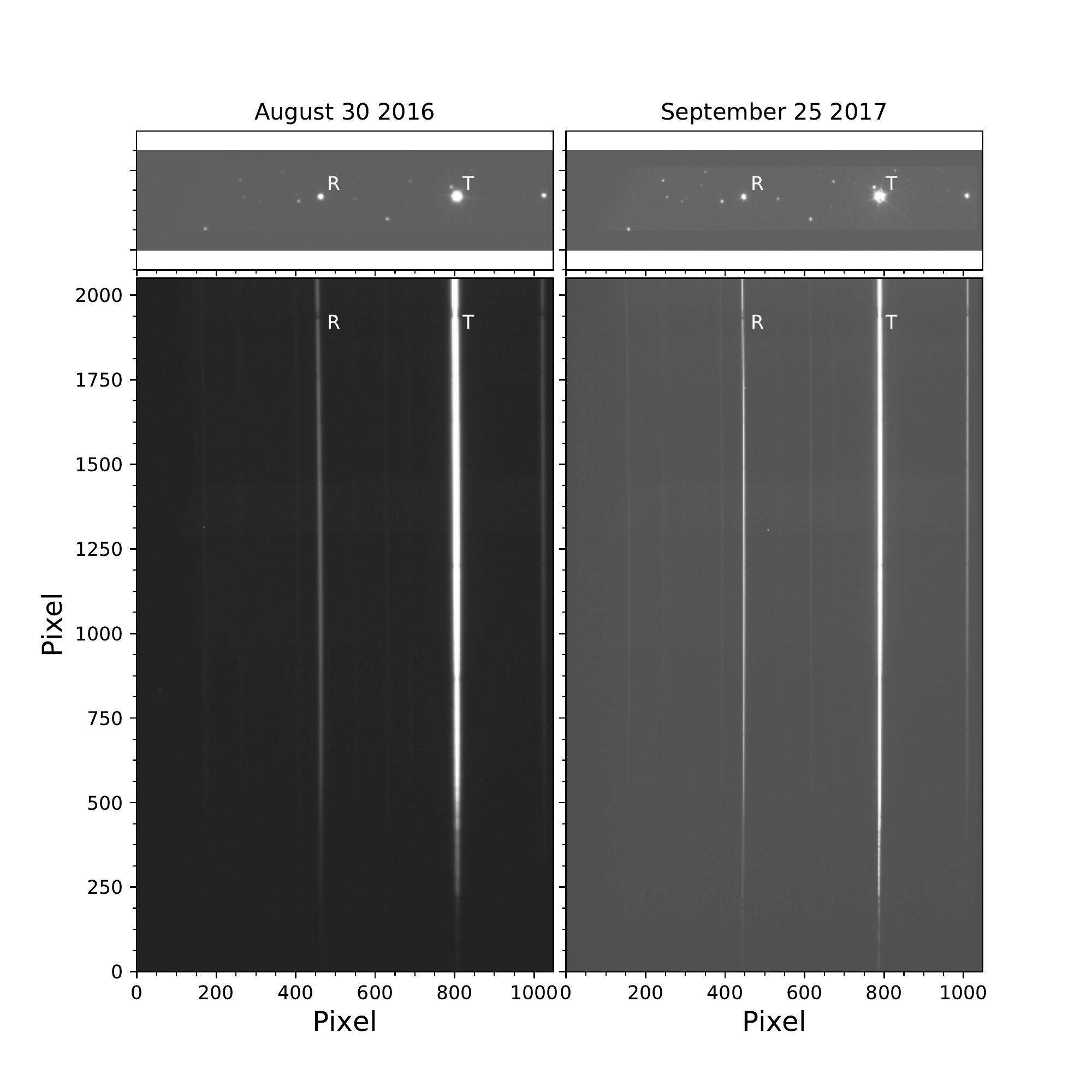}
  \caption{GTC/OSIRIS image through the slit (top) and raw science image (bottom) for the nights of August 30 2016 (left) and September 25 2017 (right). In this image, the target star is labeled T and the selected reference star is labeled R.}
  \label{Fig:CCD}
\end{figure}

\begin{figure}
  \centering
  \includegraphics[width=\hsize]{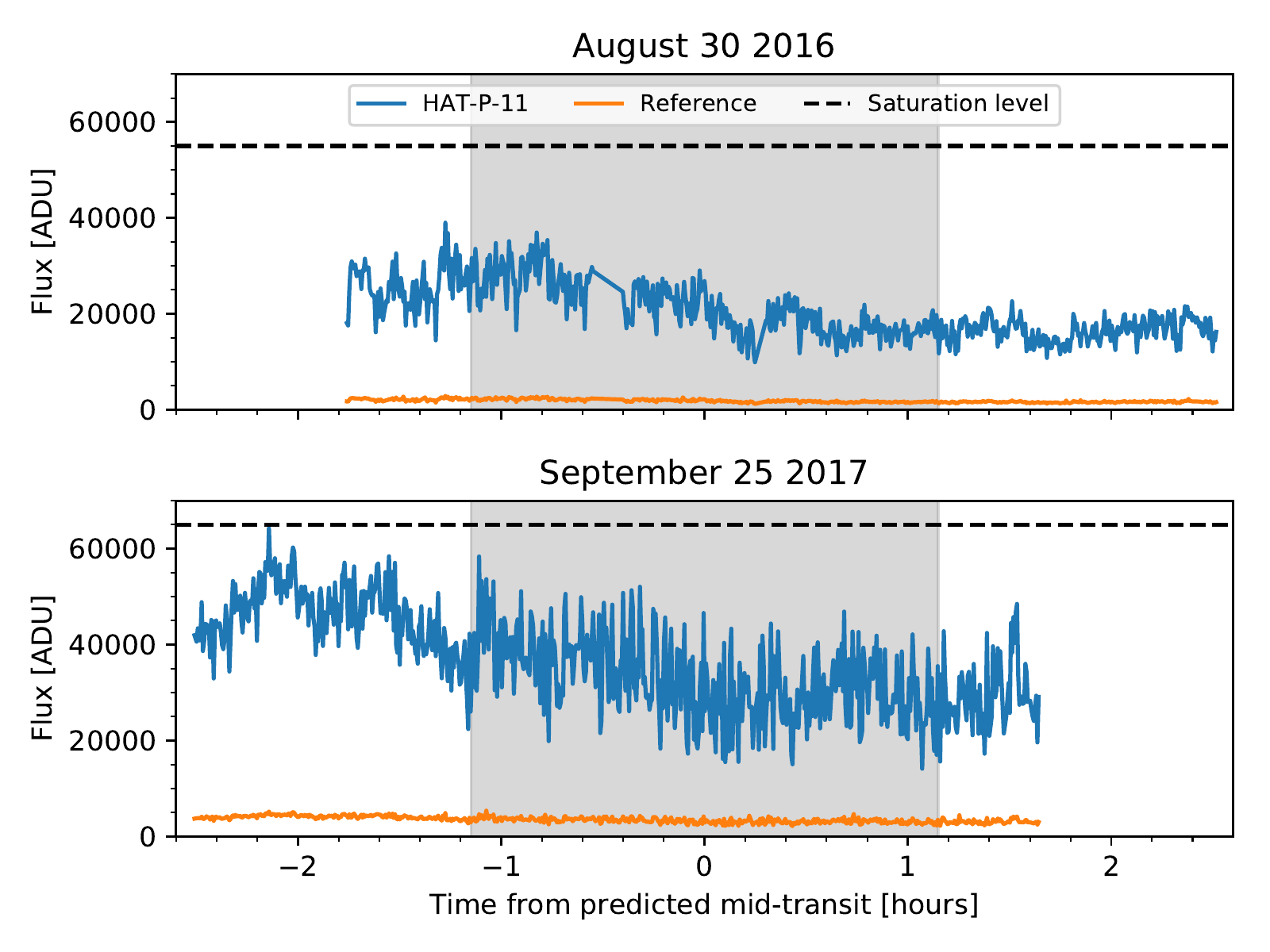}
  \caption{HAT-P-11 and reference star count level of the pixel with the maximum flux inside the extraction aperture for the nights of August 30 2016 (top) and September 25 2017 (bottom). The shaded area marks the duration of the transit and the dashed line marks the saturation levels of the detector. The different saturation levels in the two nights are due to different readout modes adopted.}
  \label{Fig:PeakCounts}
\end{figure}

Two transits of HAT-P-11b were observed with OSIRIS in the nights of August 30 2016 (hereafter N1) and September 25 2017 (hereafter N2). For both observations we were able to put HAT-P-11 and one reference star inside the slit (see Table \ref{Table:StarDescr} for the star properties and Figure \ref{Fig:CCD} for the acquisition image and an example of a raw science frame for each night). The data were collected using the R1000B grism ($\lambda \sim$ 350 - 787 nm of coverage) and a custom built slit of 40 arcsec of width. The instrumental configuration for both nights is described in Table \ref{Table:InstConf}. For the first night the exposure time was set to 4.25 seconds and the total time of observation was 4.27 hours (from 20:43 UT to 01:00 UT) which translated into 693 science images. At 21:57 UT we lost 9 minutes of data due to problems with the telescope software, after the problem was solved the stars were reacquired at the same initial position inside the slit. For the night of September 25 2017 the exposure time was set to 3.50 seconds and the total time of observation was 5.07 hours (from 20:14 UT to 01:16 UT) which translated into 803 science images. Figure \ref{Fig:PeakCounts} shows the count level for the pixel with the maximum flux inside the extraction aperture for each star and night\footnote{saturation levels taken from \url{http://www.gtc.iac.es/instruments/osiris/media/OSIRIS-USER-MANUAL\_v3\_1.pdf}}.

\begin{table*}
  \caption{Observed stars with GTC.}
  \label{Table:StarDescr}  
  \centering                          
  \begin{tabular}{ccccccc}        
    \hline\hline                 
    Star & 2MASS name & RA (FK5 J2000) & DEC (FK5 J2000) & B (mag)  & V (mag) & K (mag)\\    
    \hline                
    HAT-P-11 & J19505021+4804508 & 19$^h$ 50$^m$ 50.21$^s$ & +48$^\circ$ 04$^m$ 50.85$^s$ & 10.66$^{(1)}$ & 9.47$^{(1)}$ & 7.01$^{(3)}$ \\
    Reference & J19505813+4805365 & 19$^h$ 50$^m$ 58.13$^s$ & +48$^\circ$ 05$^m$ 36.51$^s$ & 12.95$^{(2)}$ & 12.25$^{(2)}$ & 10.61$^{(3)}$ \\
    \hline                                   
  \end{tabular}

  \tablebib{(1)~\citet{Hog2000}; (2) \citet{Henden2015}; (3) \citet{Cutri2003} }    
\end{table*}

\begin{table*}
  \caption{GTC instrumental configuration used in both observing nights.}             
  \label{Table:InstConf}      
  \centering                          
  \begin{tabular}{c c c c c c c c}        
    \hline\hline                 
    Date & Grism &   $\lambda$ range & Binning & Readout speed & Gain & Readout noise & Slit width \\    
    \hline                        
    August 30 2016 & R1000B & 360 - 780 nm  & $2\times 2$  & 500 KHz & 1.46 $e^-$/ADU & 8.0 $e^-$ & 40 arcsec \\      
    September 25 2017 & R1000B & 360 - 780 nm  & $2\times 2$ & 200 KHz & 0.95 $e^-$/ADU & 4.5 $e^-$ & 40 arcsec \\
    
    \hline                                   
  \end{tabular}
\end{table*}

The data were reduced using the same approach described in \cite{Chen2017}. The science images were calibrated using a standard procedure (bias and flat correction), the wavelength correction was done using the 2D arc lamp images to create a pixel-to-wavelength transformation map. The spectra were extracted using an optimal extraction algorithm (\citealp{Horne1986}). A fixed extraction aperture size of 42 and 37 pixels were adopted for N1 and N2 respectively. The aperture size selected for the extraction was applied to both target and reference star and is the one that delivered the lowest scatter in the points out of transit of the white light curve, i.e. the curve produced by integrating the flux of both target and reference star in the $\lambda$ 399-785 nm region (excluding the region near the O$_2$ telluric line, $\lambda$ 757-768 nm). The mid exposure time was computed using the mid point of the UT values of the opening and closening of the shutter, these mid exposure times were transformed to Barycentric Dynamical Time with the help of the \cite{Eastman2010} code.

\begin{figure}
  \centering
  \includegraphics[width=\hsize]{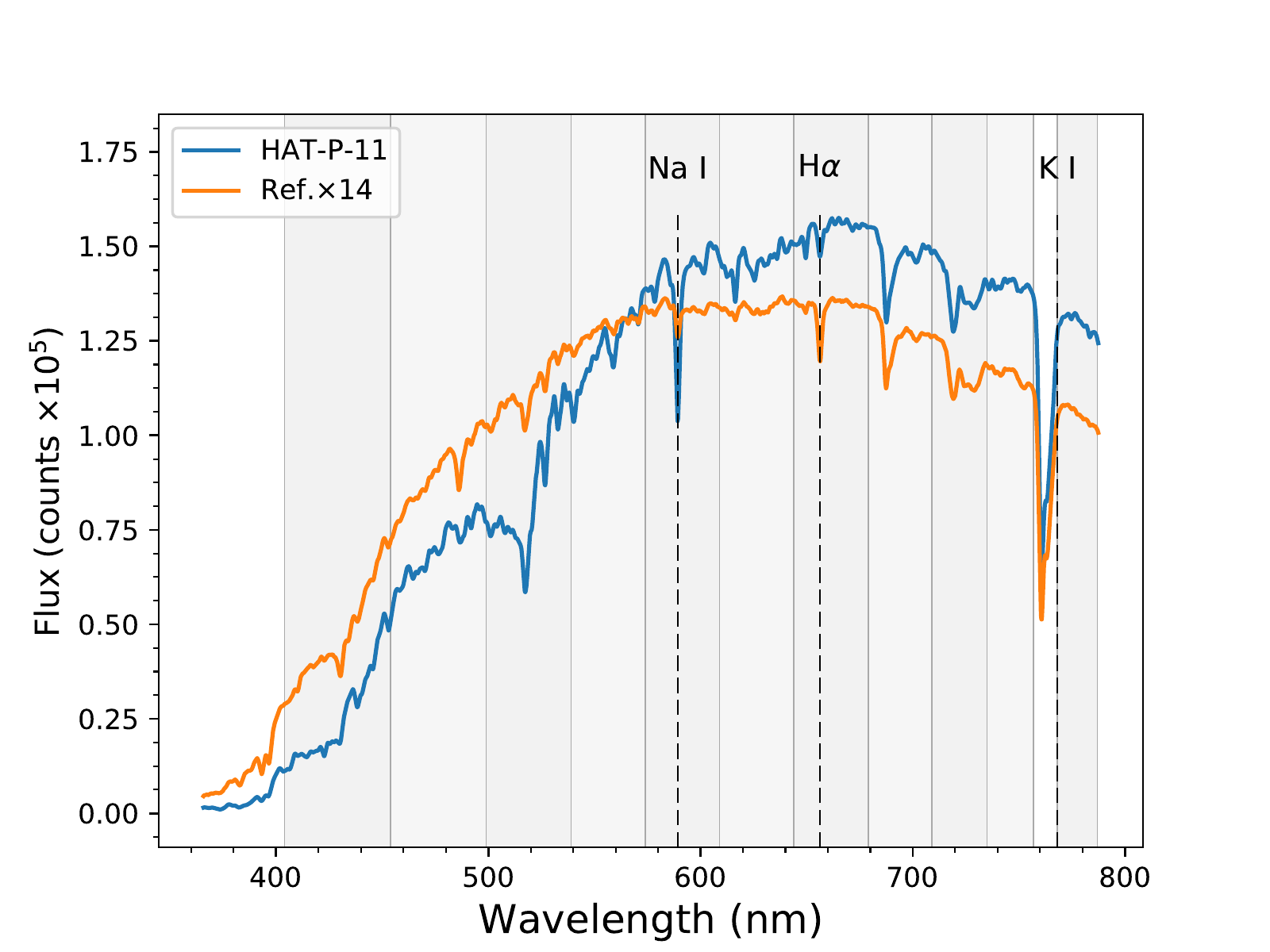}
  \caption{Extracted R1000B grism spectrum of HAT-P-11 (blue) and its reference star (orange). The spectra are not corrected for instrumental response or flux calibrated. The flux of the reference star was multiplied by a factor of 14 to match the flux level of the target for easy viewing. The shaded gray areas indicate the custom passbands used to create the spectroscopic light curves.}
  \label{Fig:StarSpec}
\end{figure}

Figure \ref{Fig:StarSpec} shows an example of the extracted spectrum of HAT-P-11 and the reference star. Note that the reference is $2.78$ mag fainter than the target in $V$ band, so the flux of the reference star is arbitrarily multiplied in the figure by a factor of 14.

\section{Data analysis}
\label{Sec:DataRed}
The data for both nights were fitted following the procedure of \cite{Chen2018}. The analytic transit curve was created using the models of \cite{MandelAgol2002} through the \texttt{batman} implementation of \cite{Kreidberg2015}. For the limb darkening coefficients (LDC) we adopted a quadratic limb darkening law; the coefficient values $u_1$ and $u_2$ were computed interpolating ATLAS (\cite{Kurucz1979} synthetic stellar spectrum with the code of \cite{Espinoza2015}. The coefficients were estimated using the stellar parameters of HAT-P-11 presented in \cite{Bakos2010} ($T_{eff}=4750$ K, [Fe/H] $=0.3$ dex, and $\log g_\star \; (cgs)= 4.5$). To model the white light curve transit we set as free parameters: the planet-to-star radius ratio $R_\mathrm{p}/R_\mathrm{s}$, the quadratic limb darkening coefficients $u_1$ and $u_2$, the central time of transit $T_\mathrm{mid}$ (for N1 and N2), the orbital semi-major axis over stellar radius $a/R_\mathrm{s}$, and the orbital inclination $i$. For the spectroscopic light curves the orbital parameters $i$ and $a/R_\mathrm{s}$ were fixed to the values found for the white light curve. The orbital eccentricity $e$, period, and argument of the periastron $\omega$ were fixed to the values presented in \cite{Huber2017} for all the curves analyzed here.

We created a common-mode noise model by dividing the white light curve by the best fitting transit model, and then each spectroscopic light curve was divided by this common-mode noise. To model the systematic effects present in the data, we used the Gaussian Processes (GP) python implementation provided by \texttt{george} (\citealp{Ambikasaran2015}). The use of GP to model the red noise has become a standard procedure in light curve analysis (e.g., \citealp{Gibson2012}). For all the red noise sources, we selected a squared exponential kernel described by

\begin{equation}
  k(x_i,x_j) = A^2 \exp \left[ -\sum^{N}_{\alpha=1} \left( \frac{x_{\alpha,i}-x_{\alpha,j}}{L_\alpha} \right)^2 \right]
\label{Eq:SEKernel}  
\end{equation}
where $x_{i,j}$ are the input vectors associated with the origin of the red noise, in this case a time component and seeing variation measured by the full width at half maximum (FWHM) of the spectral profile in the spatial direction of the target.

\begin{figure}
  \centering
  \includegraphics[width=\hsize]{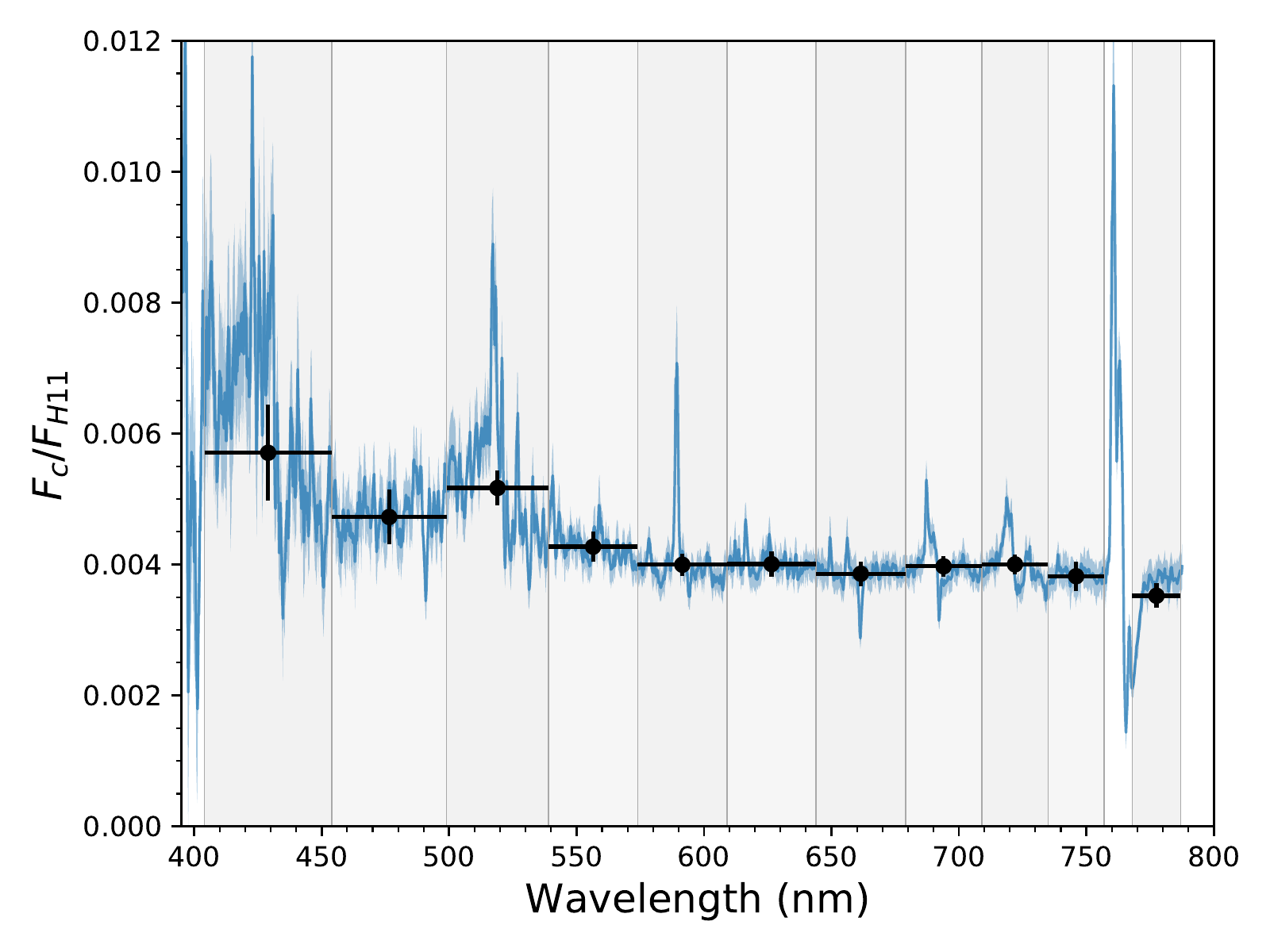}
  \caption{Flux ratio between HAT-P-11 and the contaminant star (blue line). The black dots show the mean flux contamination level inside the 12 bins used to compute HAT-P-11b transmission spectrum. The flux contamination level is relatively stable across the probed wavelength range.}
  \label{Fig:StarContamination}
\end{figure}

Near the position of the target there is a faint star whose flux is inside the spectral extraction aperture for both nights. GTC was able to observe a transit of HAT-P-11b in August 12 2017, unfortunately the target was saturated and in non-linearity range in most of the in-transit images so we decided to not use this data set in the transmission spectroscopy analysis. The seeing on that night was stable enough to allow us to separate the flux of HAT-P-11 and the contaminant star (see Figure \ref{Fig:StarContamination}). The flux of the contaminating star was extracted from the non-saturated pre-transit frames of the August 12 2017 data set (see Appendix). We included the flux contamination level by this contaminant in the MCMC procedure to compute an accurate planet-to-star radius ratio ($R_\mathrm{p}/R_\mathrm{s}$) across the probed wavelength range. 

A likelihood function was evaluated iteratively using the python Markov chain Monte Carlo (MCMC) suite \texttt{emcee} (\citealp{ForemanMackey2013}). For the white light curve we used normal priors for the following parameters: the LDC with values centered around the predicted coefficients and with a normal distribution width of 0.1 plus the LDC constrains by \cite{Kipping2013}, the orbital inclination and semi-major axis over stellar radius (mean centered around \citealp{Huber2017} values), and the flux contamination fraction of the contaminant star. For the spectroscopic light curves we used normal priors for the LDC and the flux contamination fraction of the contaminant star. For the rest of the parameters we used uniform priors for both white light and spectroscopic curves.

The MCMC for the white light curve consisted of 90 independent chains (or walkers) and run for 3000 iterations. For the spectroscopic curves we used 32 independent chains and 3000 iterations. After discarding the first 500 iterations as burn-in period, we computed the median and 1$\sigma$ values of the posterior distribution for each parameter.

\section{Results and discussion}
\label{Sec:Disc}

\subsection{White light curve}
\begin{figure}
  \centering
  \includegraphics[width=\hsize]{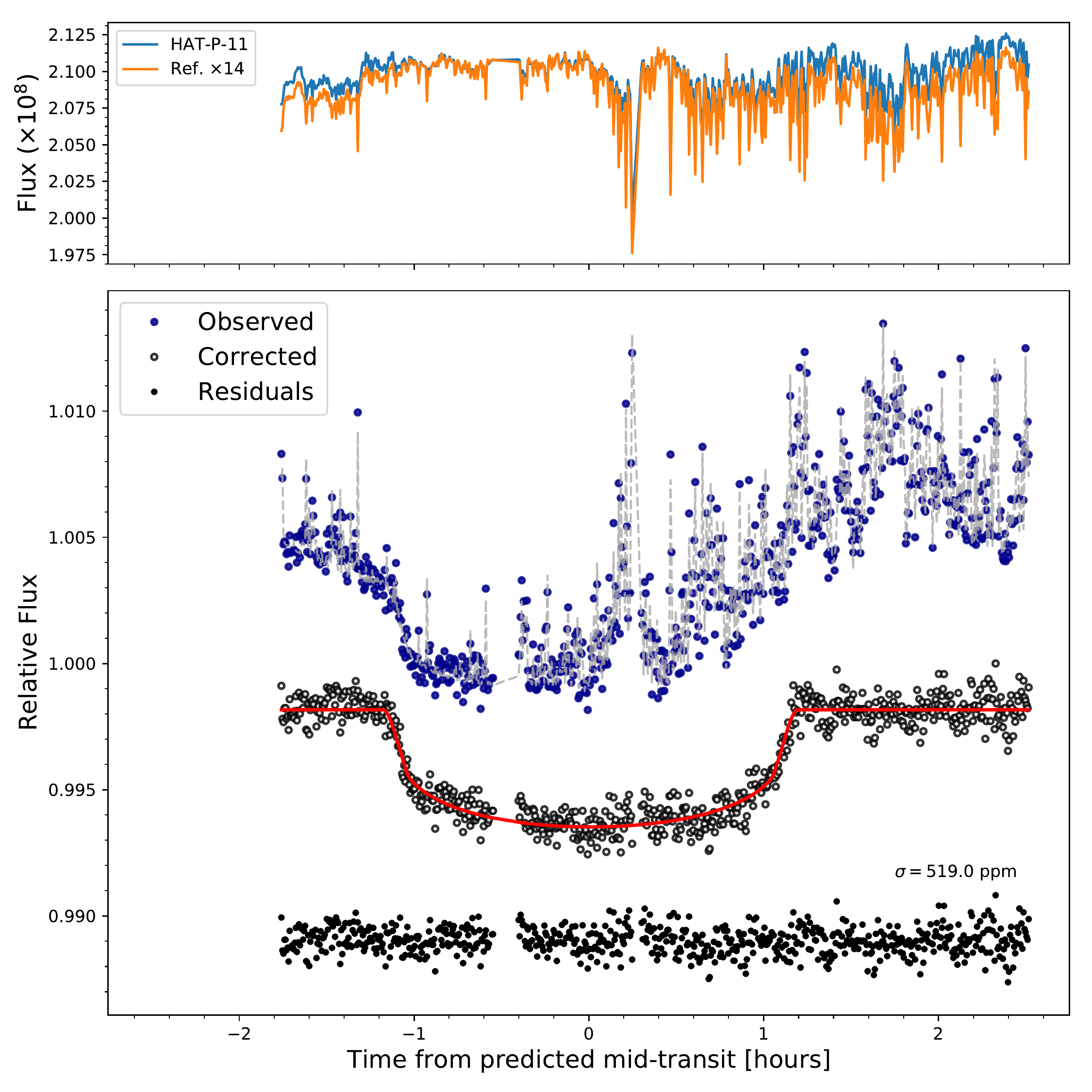}
  \caption{GTC/OSIRIS HAT-P-11b white light transit curve for the night of August 30 2016. \textit{Top panel:} Measured flux vs time of HAT-P-11 and the reference star; the flux of the reference was multiplied by a factor of 14 for displaying purposes. \textit{Bottom panel:} The blue points represent the observed time series, the gray dashed line shows the best-fit model (transit model plus systematics and red noise) determined using our MCMC analysis, the black open circles show the light curve after removing the systematic effects and red noise component while the red line is the best-fitting transit model. All the curves are shown with an arbitrary offset in y-axis.}
  \label{Fig:WLCurve_N1}
\end{figure}

\begin{figure}
  \centering
  \includegraphics[width=\hsize]{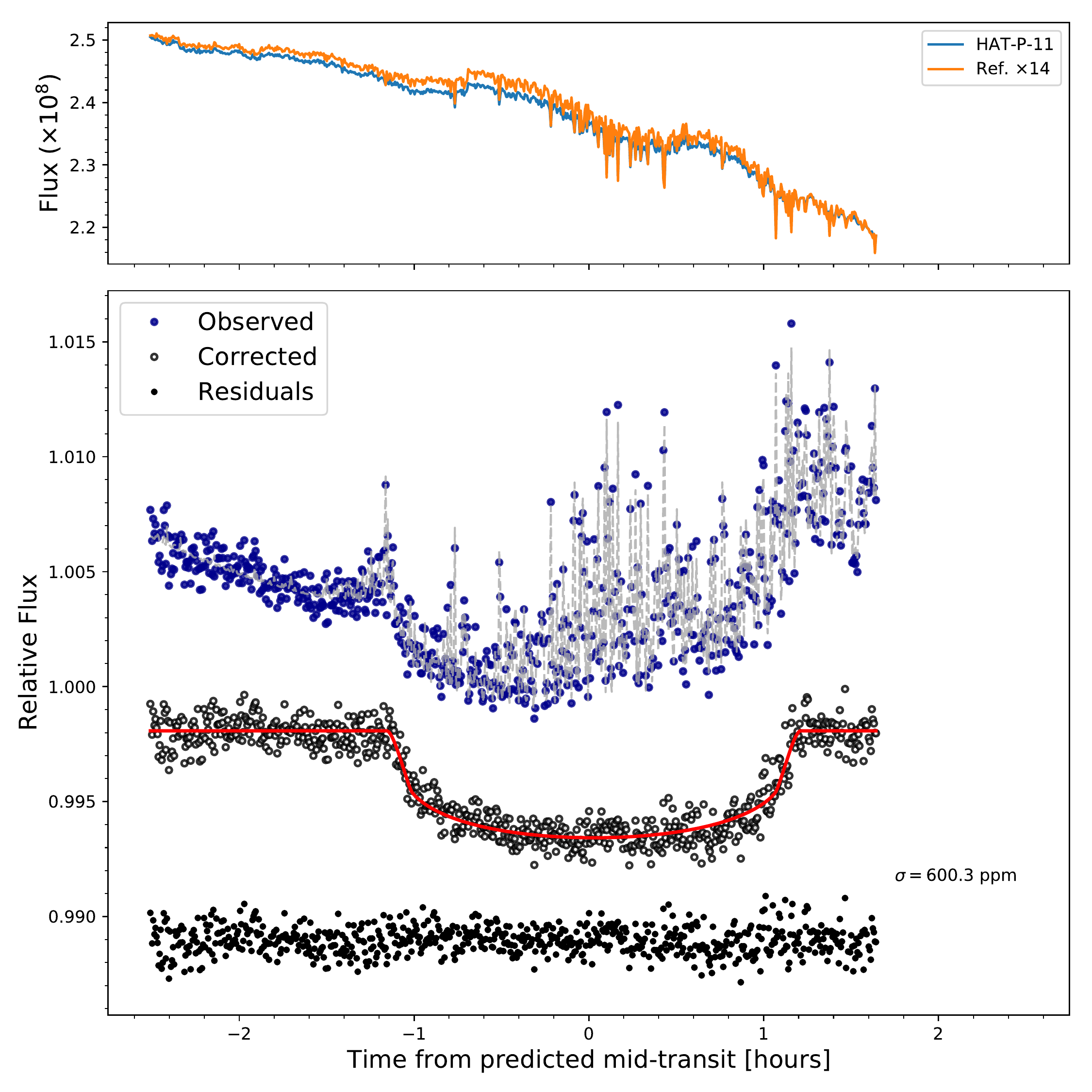}
  \caption{Same as Figure \ref{Fig:WLCurve_N1} for the night of September 25 2017.}
  \label{Fig:WLCurve_N2}
\end{figure}

The white light curves for N1 and N2 are shown in Figures \ref{Fig:WLCurve_N1} and \ref{Fig:WLCurve_N2}. Since the reference star is $\sim 2.7$ magnitudes fainter than HAT-P-11 in $V$ band, the reference star was more affected by seeing variations during the observations and it is the major source of point-to-point variation in the light curve. The seeing variations affect the number of pixels where the point spread function (PSF) is contained and this effect is correlated with the FWHM of the stellar flux. Our light curve modeling is able to reproduce this noise source and we can recover the transit for each night. Once we subtract the transit model from the light curves, we reach a root mean square (RMS) of the residuals of 519 ppm and 600 ppm for N1 and N2 respectively. The transit parameters and 1$\sigma$ uncertainties fitted using a joint analysis of N1 and N2 are listed in Table \ref{Table:WL}.

\begin{table}
  \caption{MCMC joint analysis results of the white light transit curve of HAT-P-11b.}
  \label{Table:WL}
  \centering
  \begin{tabular}{lr}
    
    \hline\hline
    Transit parameter & Value \\
    \hline
    $R_\mathrm{p}/R_\mathrm{s}$ & $0.06118 \pm 0.00181$ \\
    $u_1$ & $0.543 \pm 0.080$ \\
    $u_2$ & $0.145 \pm 0.079$ \\
    $T_\mathrm{mid}$ [BJD$_\mathrm{TBD}$] & $2457631.440482 \pm 0.00021$ \\
    $T_\mathrm{mid}$ [BJD$_\mathrm{TBD}$] & $2458022.465369 \pm 0.00024$ \\
    $a/R_\mathrm{s}$ & $14.638 \pm 0.082$ \\
    $i$ [deg] & $88.997 \pm 0.132$ \\
    $\omega$ [deg] & $-162.149^{(1)}$ (fixed) \\
    Period [days] & $4.887802443^{(1)}$ (fixed) \\
    $e$ & $0.26493^{(1)}$ (fixed) \\
    \hline
    GP parameter & Value \\
    \hline
    $\ln(A)$ N1 & $-5.323 \pm 0.281$ \\
    $\ln(L_{t})$ N1 & $-3.058 \pm 0.117$ \\
    $\ln(L_{fwhm\;y})$ N1 & $0.937 \pm 0.211$ \\
    $\ln(A)$ N2 & $-5.245 \pm 0.401$ \\
    $\ln(L_{t})$ N2 & $-2.398 \pm 0.346$ \\
    $\ln(L_{fwhm\;y})$ N2 & $1.452 \pm 0.211$ \\
    \hline
                                 
  \end{tabular}
  \tablebib{(1)~\citet{Huber2017}}
\end{table}

\subsection{Transmission spectrum}
\begin{figure}
  \centering
  \includegraphics[width=\hsize]{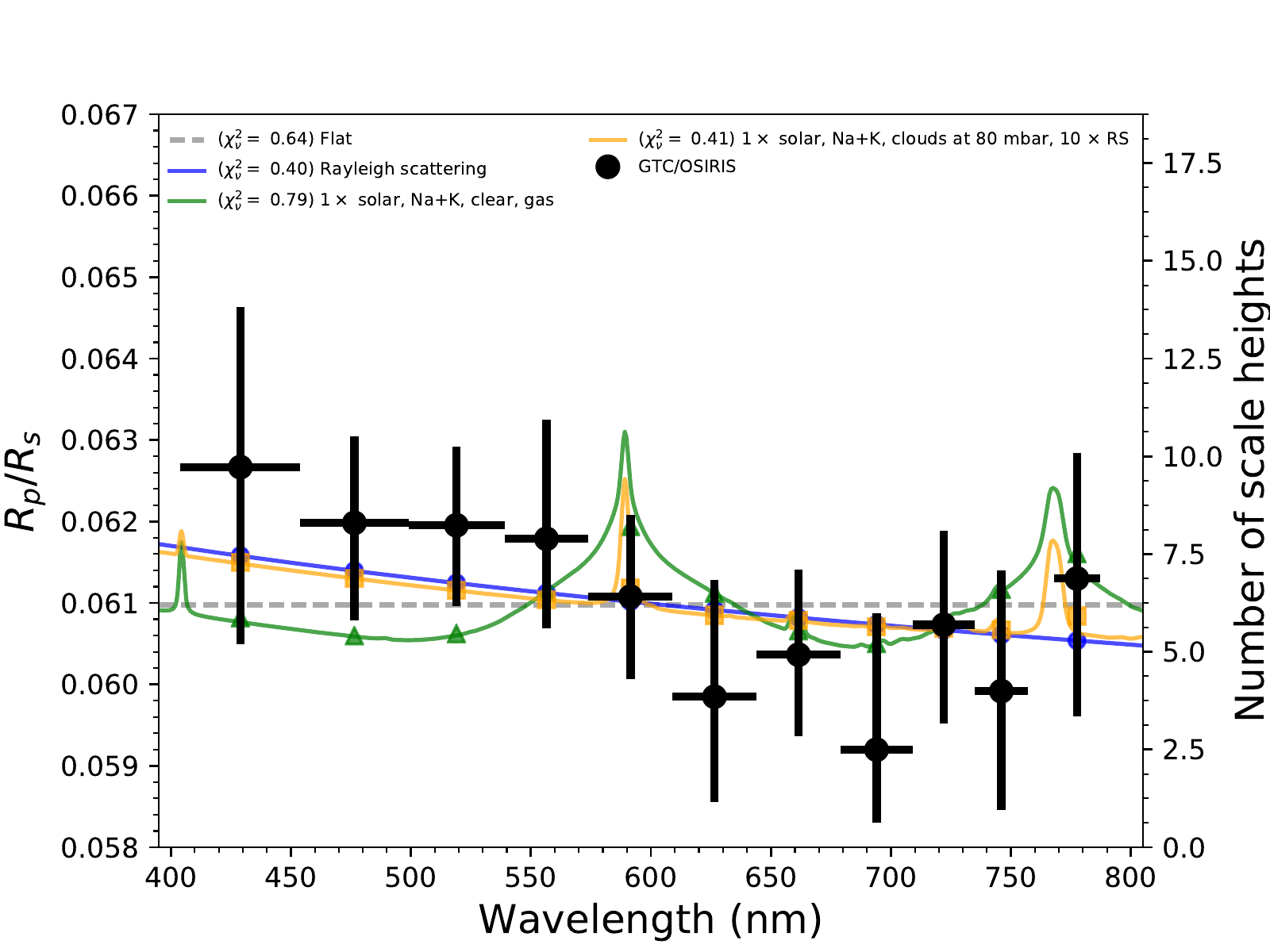}
  \caption{Optical transmission spectrum of HAT-P-11b taken with GTC/OSIRIS for the joint fit of N1 and N2.}
  \label{Fig:GTC_TransSpec}
\end{figure}

The GTC/OSIRIS transmission spectrum for the joint fit of N1 and N2 is shown in Figure \ref{Fig:GTC_TransSpec}. Our joint fit results are consistent with a pure Rayleigh scattering feature when compared to a flat model (straight line) and \texttt{Exo-Transmit} (\citealp{Kempton2016}) models with clear (i.e. no clouds) and cloudy atmospheres. The cross-section of the particles that produce the Rayleigh scattering follow a power law $\sigma = \sigma_0(\lambda / \lambda_0)^\alpha$, where $\lambda_0$ is the cross-section size at the reference wavelength $\lambda_0$ and $\alpha=-4$ (\citealp{LecavelierDesEtangs2008}). The change of measured planet-to-star radius ratio across wavelength is related to the planetary atmospheric scale height and $\alpha$ by 

\begin{equation}
  \frac{d R_\mathrm{p}}{d \ln \lambda} = \alpha H = \alpha \frac{\kappa_\mathrm{B} T_\mathrm{p}}{\mu_\mathrm{m} g_\mathrm{p}}
  \label{Eq:RayleighScattering}
\end{equation}
where $H$ is the atmospheric scale height, $\kappa_\mathrm{B}$ the Boltzmann constant, $T_\mathrm{p}$ the temperature of the atmosphere of the planet, $\mu_\mathrm{m}$ the mean molecular weight of the planetary atmosphere, and $g_\mathrm{p}$ is the planet's surface gravity.

Assuming a hydrogen-dominated atmosphere ($\mu_\mathrm{m}=2.37$), a planetary temperature of $T_\mathrm{p} = 878 \pm 15 $ K (\citealp{Bakos2010}), and adopting $g_\mathrm{p}=14.08 \pm 1.17\; m/s^2$ (computed using the planetary mass of \citealp{Southworth2011} and radius of \citealp{Deming2011}), we find $\alpha_{joint} = -9.86 \pm 6.19 $ for the measured $d R_\mathrm{p} / d \ln \lambda$ using the joint fit of N1 and N2. It is worth mentioning that the relatively large uncertainty in $\alpha_{joint}$ comes from the stellar radius of HAT-P-11 (eq. \ref{Eq:RayleighScattering} needs the planetary radius, not the planet-to-star radius ratio). Considering the uncertainties, the $\alpha$ value for the joint fit is consistent with Rayleigh scattering from a hydrogen dominated atmosphere, although the measured value departs from the expected $\alpha=-4$. This led us to examine in closer detail each observing night.

\begin{figure}
  \centering
  \includegraphics[width=\hsize]{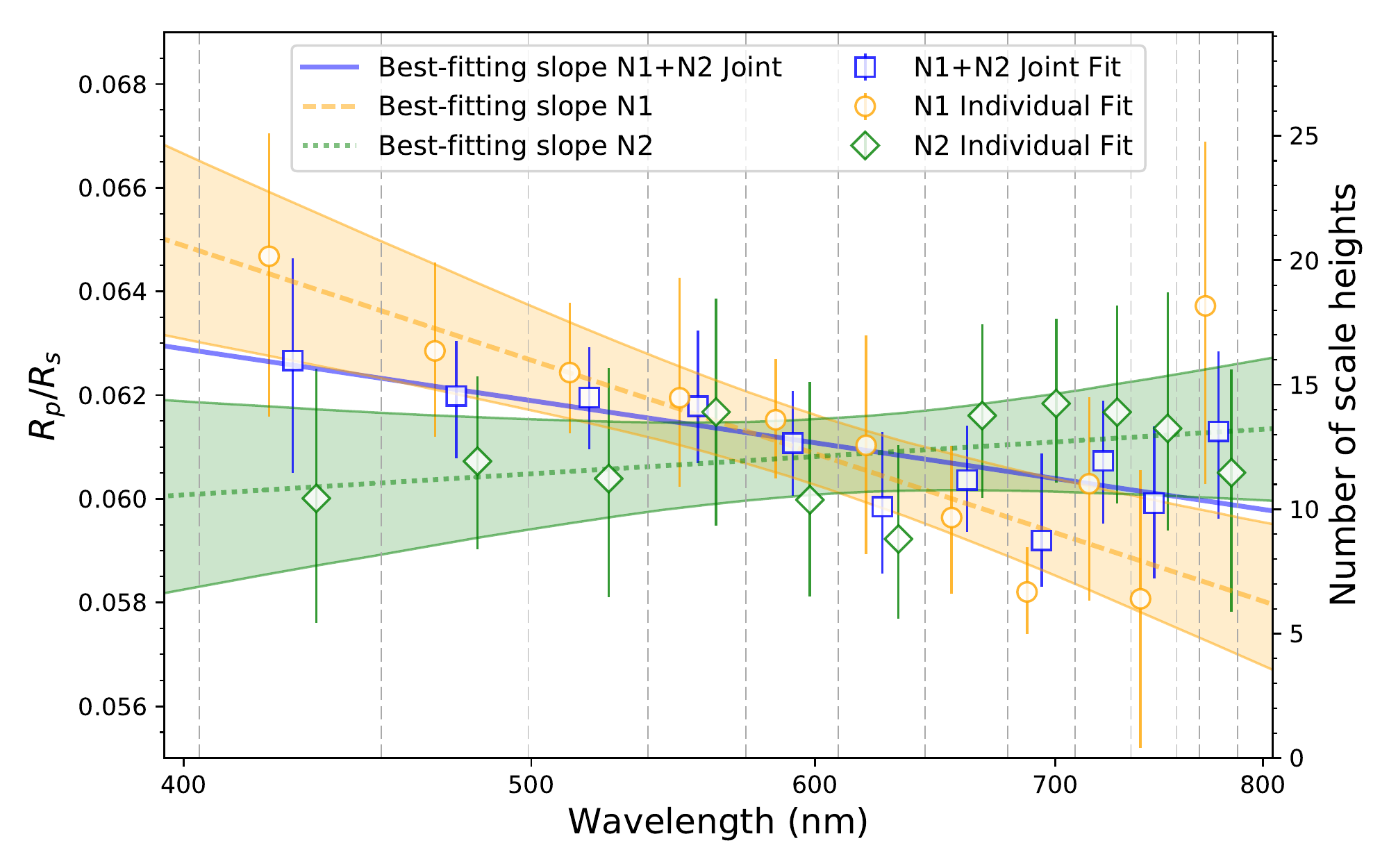}
  \caption{Optical transmission spectrum of HAT-P-11b taken with GTC/OSIRIS for the joint fit (blue), and individual fit for N1 (orange) and N2 (green). The lines and shaded areas represent the best fit and 1$\sigma$ uncertainty range for N1 (orange) and N2 (green). The data points are shifted in x-axis arbitrarily.}
  \label{Fig:RS_fit}
\end{figure}

We re-analyzed the transits for N1 and N2, this time leaving the planet-to-star radius ratio as free parameter for each night independently. Figure \ref{Fig:RS_fit} presents a comparison of the transmission spectrum of HAT-P-11b for each night fitted independently along with the results of the joint fit. The data from N1 presents a greater $R_\mathrm{p}/R_\mathrm{s}$ versus $\ln \lambda$ slope; if we use the same $\mu_\mathrm{m}$, $T_\mathrm{p}$, and $g_\mathrm{p}$ values as before we get $\alpha_{N1} = -21.85 \pm 9.25$. In contrast the transmission spectrum for N2 is relatively flat and featureless with a slope slightly increasing towards redder wavelengths. One possible explanation for the discrepancy between data sets is stellar activity.

\begin{table*}
  \caption{Measured $R_\mathrm{p}/R_\mathrm{s}$ for HAT-P-11b.}
  \label{Table:RpRs_bins}
  \centering
  \begin{tabular}{cccccc}

    \hline\hline
    $\#$ & Center (nm) & Width (nm) & $(R_\mathrm{p}/R_\mathrm{s})$ Night 1 & $(R_\mathrm{p}/R_\mathrm{s})$ Night 2 & $(R_\mathrm{p}/R_\mathrm{s})$ Joint Fit  \\
    \hline
    1 & 429.0 & 50 & $0.06468 \pm 0.00273$ & $0.06001 \pm 0.00245$ & $0.06267 \pm 0.00207$ \\
    2 & 476.5 & 45 & $0.06285 \pm 0.00168$ & $0.06073 \pm 0.00166$ & $0.06198 \pm 0.00113$ \\
    3 & 519.0 & 40 & $0.06244 \pm 0.00126$ & $0.06039 \pm 0.00220$ & $0.06195 \pm 0.00098$ \\
    4 & 556.5 & 35 & $0.06195 \pm 0.00202$ & $0.06167 \pm 0.00219$ & $0.06179 \pm 0.00128$ \\
    5 & 591.5 & 35 & $0.06153 \pm 0.00115$ & $0.05998 \pm 0.00207$ & $0.06108 \pm 0.00101$ \\
    6 & 626.5 & 35 & $0.06103 \pm 0.00211$ & $0.05923 \pm 0.00167$ & $0.05985 \pm 0.00136$ \\
    7 & 661.5 & 35 & $0.05964 \pm 0.00143$ & $0.06161 \pm 0.00168$ & $0.06036 \pm 0.00102$ \\
    8 & 694.0 & 30 & $0.05820 \pm 0.00084$ & $0.06184 \pm 0.00158$ & $0.05920 \pm 0.00129$ \\
    9 & 722.0 & 26 & $0.06029 \pm 0.00196$ & $0.06167 \pm 0.00191$ & $0.06073 \pm 0.00118$ \\
    10 & 746.0 & 22 & $0.05807 \pm 0.00268$ & $0.06136 \pm 0.00230$ & $0.05992 \pm 0.00147$ \\
    11 & 777.5 & 19 & $0.06372 \pm 0.00331$ & $0.06051 \pm 0.00234$ & $0.06130 \pm 0.00162$ \\
    \hline
    
  \end{tabular}
\end{table*}

\subsection{The impact of stellar spots}
The host star of HAT-P-11b is known to be active, in the discovery paper \cite{Bakos2010} measured a mean Ca\,{\sc ii} H \& K emission activity index of $\langle S \rangle= 0.61$. \cite{Morris2017} studied the evolution of the $S$-index of HAT-P-11 from 2008 to 2017 using Keck/HIRES data, they found evidence of an activity cycle close to 10 years of duration and measured a mean $S$-index of $\langle S \rangle= 0.58 \pm 0.04$. The mean $S$-index measured by \cite{Morris2017} is equivalent to an activity index of $\log R'_{HK} = -4.35$ (see \citealp{Noyes1984} for the definition of the $R'_{HK}$ index), a value representative for stars exhibiting significant stellar activity.

One way in which stellar activity affects the star is the appearance of spots in its surface. There are several transits of HAT-P-11b observed by \textit{Kepler} where the planet crossed regions with spots (for example see \citealp{Southworth2011} Fig. 23). \cite{SanchisOjeda2011} used \textit{Kepler} data to constrain the orbital obliquity and identified two regions with spots which were interpreted as active regions of the star. Also using \textit{Kepler} data \cite{Beky2014} found evidence of a long-lived spot region in HAT-P-11 that was eclipsed by the planet 6 times and discovered that the orbital period of HAT-P-11b and the rotational period of the host star are in a period ratio of 6:1. \cite{Morris2017a} studied the spot area coverage and spot distribution across the stellar surface, deducing that $3^{+6}_{-1}$ \% of the surface is covered by spots and that they are distributed around $16^\circ \pm 1^\circ$ latitude.

Occulted and unocculted stellar spots can affect the flux measurement during a transit, leading to a wrong estimate of the transit depth (\citealp{Pont2013}). According to \cite{Fraine2014} the median amplitude of spot crossing events in \textit{Kepler} band is close to 500 ppm, by visual inspection we did not detect any spot crossing event in both of our data sets which both have rms of the residuals of the order of 500 ppm in the white light curve. The major problem preventing us from detecting occulted spots is the faintness of the reference star, which introduced strong systematics correlated with seeing. Even if there were spot crossing events in our data sets, their effect on the measured transmission spectrum would be negligible due to the modeling of common-mode noise together with Gaussian Processes (CMN removes any abrupt change common in wavelength, GP treat the spot as time-correlated noise). However, there is still a chance that unocculted spots can be responsible of producing a signal similar to Rayleigh scattering in the transmitted spectrum of the planet for N1. 

Since spots are regions with different temperature than the average surface temperature, they can introduce a slightly different stellar flux level in some wavelengths creating an effect similar to Rayleigh scattering. To model the effect of unocculted spots during the HAT-P-11 b transit of N1 we followed the approach of prior works (e.g., \citealp{Sing2011}, \citealp{McCullough2014}, \citealp{Chen2017}, \citealp{Rackham2018}) in which the observed planet-to-star radius ratio (without including the effect of limb darkening) can be modeled by 

\begin{equation}
  \left( \frac{\bar{R_\mathrm{p}}}{\bar{R_\mathrm{s}}} \right)^2 = \left( \frac{R_\mathrm{p}}{R_\mathrm{s}} \right)^2 \frac{1}{1 - \delta (1 - F_\nu(spot)/F_\nu(phot)) }
  \label{Eq:RpRs_spots}
\end{equation}
where $\bar{R_\mathrm{p}}/\bar{R_\mathrm{s}}$ is the observed planet-to-star radius ratio, $R_\mathrm{p}/R_\mathrm{s}$ is the true planet-to-star radius ratio (i.e. not affected by spots), $\delta$ is the variable that takes into account the stellar surface area covered by spots or filling factor, $F_\nu(spot)$ is the flux from the stellar spots, and $F_\nu(phot)$ is the flux from the stellar photosphere.

To model the flux from the star (and spotted regions) we used PHOENIX stellar models (\citealp{Husser2013}) for HAT-P-11 (the model with stellar parameters of $T_{eff}=4780$ K, $\log g_\star \; (cgs)= 4.5$, [Fe/H] $=0.5$ dex) and compared it with a grid of spectra with temperatures ranging from 2700 K to 7000 K.

We performed a MCMC fitting procedure using equation \ref{Eq:RpRs_spots} applied to the transmission spectrum of N1 and N2, after removing the last data point (bin centered at 777.5 nm) which presents a different $R_\mathrm{p}/R_\mathrm{s}$ than the rest of the bins probably due to noise added by the nearby telluric O$_2$ line ($\lambda=760.5$ nm, see Fig. \ref{Fig:StarSpec}). We want to emphasize that the main noise source in this bin is the contaminant star whose wavelength is shifted due to misalignment in the slit (see Fig. \ref{Fig:StarContamination}). If there were no such a contaminant star, the majority of the telluric O$_2$ band would have been excluded from our flux integration (see Fig. \ref{Fig:StarSpec}), and the remaining effects should have been corrected by the reference star.

The MCMC was launched with \texttt{emcee} using 80 independent chains and two separate runs: a burn-in run with 10000 iterations and a main run with 25000 iterations. Using the posterior distributions to get the median values and 1$\sigma$ uncertainty range, we get for N1: $\delta_{N1} = 0.62^{+0.20}_{-0.17}$ and $T_{spot\; N1}=4351^{+184}_{-299}$ K. \cite{Morris2017a} reported a filling factor of $3^{+6}_{-1}$ \% for HAT-P-11, a value that the authors suggest as a lower limit due to the assumption made that the stellar spot coverage probed by the transit is representative of the spot coverage of the whole star and the sensitivity of their measurements. In the same article, they claim that HAT-P-11's filling factor is close to the values of the active K star OU Gem presented in \cite{ONeal2001} with numbers ranging from $\delta \leq 0.04$ up to 0.35. Our median value for the filling factor is higher than the values for OU GEM, however the lower limit of the filling factor presented here ($\delta_{N1\; low} = 0.45$) is close to the numbers presented in \cite{ONeal2001} and typical values seen in active mid-K dwarfs (e.g., \citealp{Berdyugina2005}, \citealp{Andersen2015}) and M dwarfs (e.g., \citealp{Jackson2009}, \citealp{Jackson2012}). The difference in temperature between the star and the spot is $\Delta T = T_{phot}-T_{spot\; N1}= 429^{+184}_{-299}$ K (considering a stellar model with a temperature of $T_{phot} = 4780$ K), which is almost half the temperature presented in \cite{Fraine2014} who measured a $\Delta T \approx 900$ K based on the largest star spot crossing event observed by their \textit{Kepler} and \textit{Spitzer} runs. The surface-to-spot temperature difference of the largest spot crossing event observed by \cite{Fraine2014} is consistent with values found in active K dwarfs, which usually are in the range of 1000 - 1500 K (e.g., \citealp{ONeal2004}, \citealp{Berdyugina2005}).

The relative large value and uncertainties found in $\delta$ are probably caused by two factors: 1) the uncertainties of the measured planet-to-star radius ratio allowed several combinations of filling factors and spot temperatures to reproduce the observed transmission spectrum, and 2) the filling factor and the temperature of the stellar spot (which gives $F_\nu(spot)$) are correlated, thus, producing a very wide posterior probability distribution for the $\delta$ parameter.

The second data set could also be affected by bright and dark unocculted spots whose effects cancel out producing the measured flat transmission spectrum (see Fig. \ref{Fig:RS_fit}). It is possible to extend Eq. \ref{Eq:RpRs_spots} to include the effects of bright and dark unocculted spots (see \citealp{Rackham2018}), however due to our uncertainties and the strong correlation between the filling factor and temperature of the spot we decided to use the same MCMC procedure applied for N1 for simplicity. For the flat transmission spectrum of N2 we get $\delta_{N2} = 0.19^{+0.30}_{-0.14}$ and $T_{spot\; N2}=4724^{+422}_{-1488}$ K, a lower filling factor compared to N1 and a spot temperature closer to the adopted temperature of the star.

\begin{figure}
  \centering
  \includegraphics[width=\hsize]{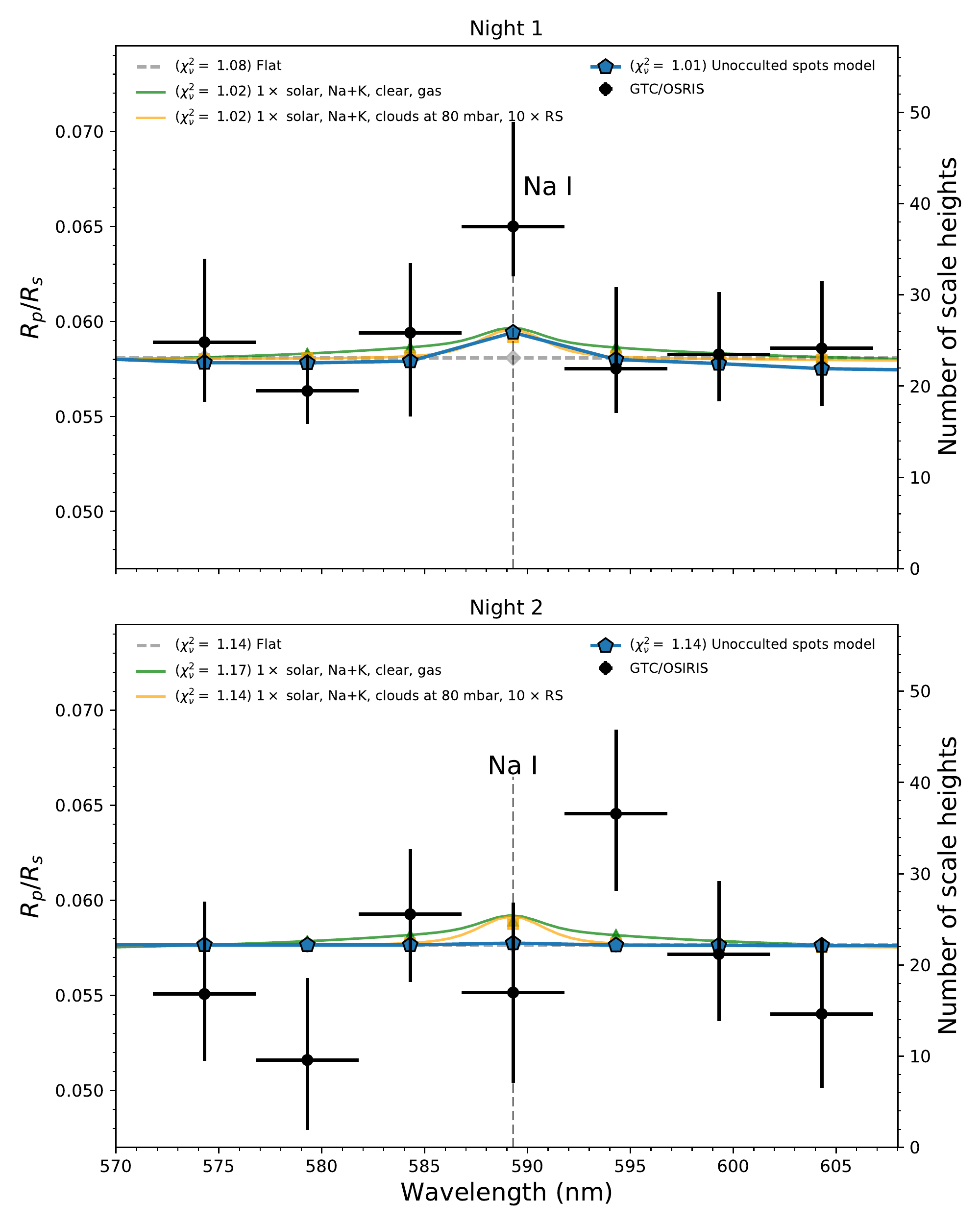}
  \caption{Transmission spectrum of HAT-P-11b around the  Na\,{\sc i} 589.0 and 589.6 nm doublet for the individual fit of N1 (top panel) and N2 (bottom panel).}
  \label{Fig:Na_Comparison}
\end{figure}

As a final test we used equation \ref{Eq:RpRs_spots} to compute the filling factor and spot temperature necessary to go from the transmission spectrum with a slope of N1 to the featureless transmission spectrum of N2. Using the same MCMC fitting procedure as before and without any constrains in the temperature of the spot we get $\delta = 0.042^{+0.278}_{-0.036}$ and $T_{spot}=4802^{+1073}_{-891}$ K. If we put the constraint that the spot temperature must be higher than the average stellar surface temperature (i.e., bright spots only), we get $\delta = 0.045^{+0.317}_{-0.041}$ and $T_{spot}=5078^{+2016}_{-263}$ K. Both results are consistent with each other considering the uncertainties, although the second test preferred higher spot temperatures ($\Delta T = 298^{+2016}_{-263}$ K) compared to the unconstrained temperature test ($\Delta T = 22^{+1073}_{-891}$ K). The low filling factor values obtained in these tests indicate that a relatively small amount of bright spots can suppress the slope found in N1 producing the flat transmission spectrum of N2.

Another way to establish if unocculted spots affected the transmission spectrum is to check the measured transit depth around some atomic lines. For example spots can create an excess in the transit depth at the Na\,{\sc i} 589.0 and 589.6 nm doublet while the transit depth decreases in H$_\alpha$ (656.3 nm) (see \citealp{Chen2017b}). We fitted (individually for N1 and N2) several light curves with a bin size of 5 nm of width around the Na\,{\sc i} doublet, but this time using the mean FWHM of the absorption lines as input for the GP instead of the FWHM of the spectral profile used in the previous presented results. The reason for this change is that using the FWHM of the absorption lines delivers better results when modeling the red noise in curves produced with narrow wavelength bins.

Figure \ref{Fig:Na_Comparison} shows the planet-to-star radius ratio measured around the Na\,{\sc i} doublet for N1 (top panel) and N2 (bottom panel), compared to the same \texttt{Exo-Transmit} models presented in Fig. \ref{Fig:GTC_TransSpec} and the expected $R_\mathrm{p}/R_\mathrm{s}$ using the best fitting model of Eq. \ref{Eq:RpRs_spots} for N1 and N2. The first dataset (N1) presents an excess in the transit depth for the bin centered at the Na\,{\sc i} doublet in agreement with the atmospheric models for HAT-P-11b and the unocculted spot model. In the second data set (N2) the excess in $R_\mathrm{p}/R_\mathrm{s}$ at the bin centered at the Na\,{\sc i} doublet disappears, and the overall transmission spectrum is more consistent with a flat model, as predicted by the unocculted spot model for N2.

\begin{figure}
  \centering
  \includegraphics[width=\hsize]{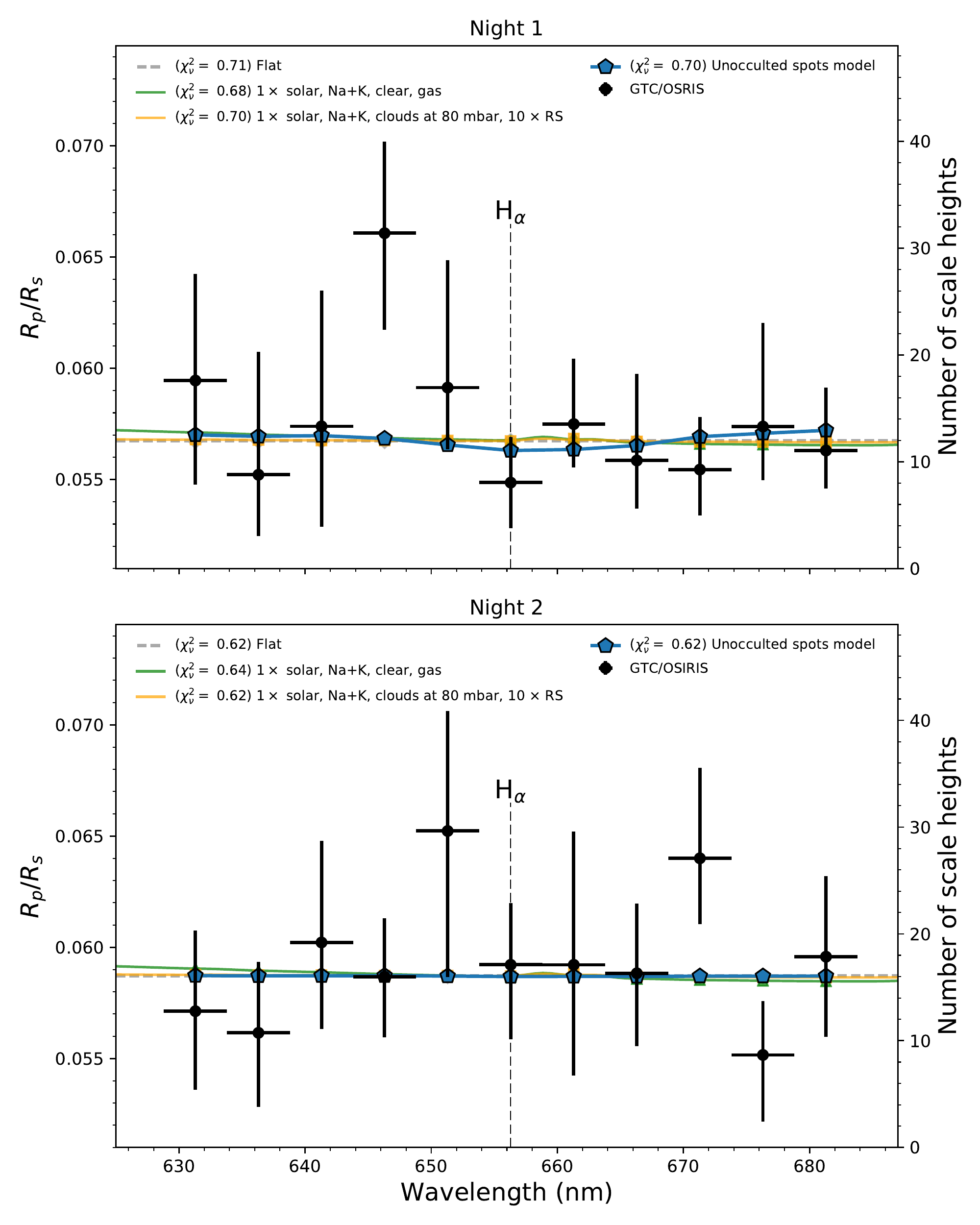}
  \caption{Transmission spectrum of HAT-P-11b around H$_\alpha$ (656.3) nm for the individual fit of N1 (top panel) and N2 (bottom panel).}
  \label{Fig:Halpha_Comparison}
\end{figure}

In the case of H$_\alpha$ our measurements are not precise enough to detect the decrease in $R_\mathrm{p}/R_\mathrm{s}$ around this line and differentiate it from a flat model (see Fig. \ref{Fig:Halpha_Comparison}).

Based on the transmission spectrum slope, unocculted spot modeling, and the transit depth excess detected in Na\,{\sc i} in N1 and not in N2, we are confident that the measured transmission spectrum of N1 was affected by stellar spots mimicking a feature similar to Rayleigh scattering.

\section{Conclusions}
\label{Sec:Conclusions}
We present here an analysis of two primary transits of the Neptune-size exoplanet HAT-P-11b taken with GTC/OSIRIS instrument. We were able to obtain long-slit spectra of HAT-P-11 and one reference star for the nights of August 30 2016 and September 25 2017. Integrating the flux of both stars in different passbands, we created several light curves to measure the change in transit depth across the wavelength range of $\lambda$ 400-785 nm. The light curves for both nights were fitted jointly and individually using a Bayesian MCMC procedure that takes into account the systematic effects present in the data.

The GTC/OSIRIS transmission spectrum of HAT-P-11b obtained with a joint MCMC fit presents a slope towards blue wavelengths. The joint fit results are consistent with a pure Rayleigh scattering feature when compared to a flat model (straight line) and synthetic atmosphere models with clear and cloudy atmospheres. Assuming that this slope was caused by small particles present in the atmosphere of the planet and adopting a planetary equilibrium temperature of $T_\mathrm{p} = 878 \pm 15 $ K and a molecular weight of $\mu_\mathrm{m}=2.37$, we measured $\alpha_{joint} = -9.86 \pm 6.19$. Considering uncertainties this $\alpha_{joint}$ value is in the expected Rayleigh scattering range caused by hydrogen-dominated atmospheres ($\alpha = -4$), although strong assumptions were made when computing this parameter (temperature of the planet and molecular weight). A closer inspection to each individual night revealed that the transmission spectrum for HAT-P-11b taken on August 30 2016 has a significant slope when compared to the transmission spectrum on the night of September 25 2017 which is mostly featureless and flat. 

The difference in the observed transmission spectrum between the two data sets could be attributed to stellar activity, more specifically to a greater number of unocculted stellar spots during N1. When modeling the amount of stellar surface area covered by dark spots and their temperature to reproduce the results of the August 2016 data set, we find $\delta_{N1} = 0.62^{+0.20}_{-0.17}$, $T_{spot\; N1}=4351^{+184}_{-299}$ K and $\Delta T = T_{star} - T_{spot\; N1} = 429^{+184}_{-299}$ K. Although our uncertainties in the filling factor are relatively large, the lower limit is in agreement with values seen in active mid-K dwarfs and M stars. The origin of the large uncertainties in the filling factor comes from the fact that $\delta$ and $T_{spot}$ are strongly correlated.

Additionally we modeled the filling factor and spot temperature needed to suppress the slope of the first data set to obtain the flat transmission spectrum of the second night for two cases: 1) setting the spot temperature completely free, and 2) forcing the spot temperature to be greater than the average stellar surface temperature (i.e., bright spots). For the first case we get $\delta = 0.042^{+0.278}_{-0.036}$ and $T_{spot}=4802^{+1073}_{-891}$ K and for the second $\delta = 0.045^{+0.317}_{-0.041}$ and $T_{spot}=5078^{+2016}_{-263}$ K. Based on these results a relative small amount of bright spots present in the star in the second night could produce the observed flat transmission spectrum, assuming that the spectrum of N1 is a closer representation to the true spectrum of the planet.

Since stellar spots can also affect the intensity of some spectral lines we also compared the transit depth around the Na\,{\sc i} doublet ($\lambda$ 589.0 nm and 589.6 nm) for both nights, finding that the transit in the first data set is deeper around the Na\,{\sc i} doublet than the nearby continuum while for the second night the transmission spectrum is flat; this is consistent with a major presence of unocculted dark spots during the transit of N1 compared to N2. The same analysis was performed for H$_\alpha$ ($\lambda$ 656.3 nm), however our transit depth measurements around this line are not precise enough to detect a change between N1 and N2.

Based on the fact that the slope in the transmission spectrum is found in N1 but not N2 and the comparison between the transit depth around Na\,{\sc i} lines for N1 and N2, we conclude that the transmission spectrum for N1 was affected by stellar activity in the form of unocculted dark spots. The second data set could also be affected by bright and dark unocculted spots whose effects cancel out producing the measured flat transmission spectrum for N2, or simply due to a low stellar activity epoch.

In light of our results for HAT-P-11b, we advocate here for repeated observations of exoplanet transmission spectroscopy studies, to account for possible stellar variability. Simultaneous/contemporary activity monitoring is crucial to establish the true nature of Rayleigh scattering features detected in exoplanets orbiting around active stars.

\begin{acknowledgements}
  Based on observations made with the Gran Telescopio Canarias (GTC), installed in the Spanish Observatorio del Roque de los Muchachos of the Instituto de Astrof\'isica de Canarias, in the island of La Palma. G.C. also acknowledges the support by the National Natural Science Foundation of China (Grant No. 11503088) and the Natural Science Foundation of Jiangsu Province (Grant No. BK20151051).\\
  \textit{Software}: \texttt{ipython} (\citealp{PER-GRA:2007}), \texttt{numpy} (\citealp{vanderwalt2011}), \texttt{scipy} (\citealp{Jones2001}), \texttt{matplotlib} (\citealp{Hunter2007}), \texttt{PyDE} (\url{https://github.com/hpparvi/PyDE}). This research made use of Astropy, a community-developed core Python package for Astronomy (\citealp{Astropy2013}, \citealp{Astropy2018}). Correlation plot for N1 and N2 spot modeling done with \texttt{Corner} (\citealp{corner}).
\end{acknowledgements}

%
%

\bibliographystyle{aa}
\bibliography{references}

\begin{appendix}
  \label{sec:appendix}

  \section{Additional figures}

  \subsection{Saturated images for August 12 2017}
\begin{figure}
    \centering
    \includegraphics[width=\hsize]{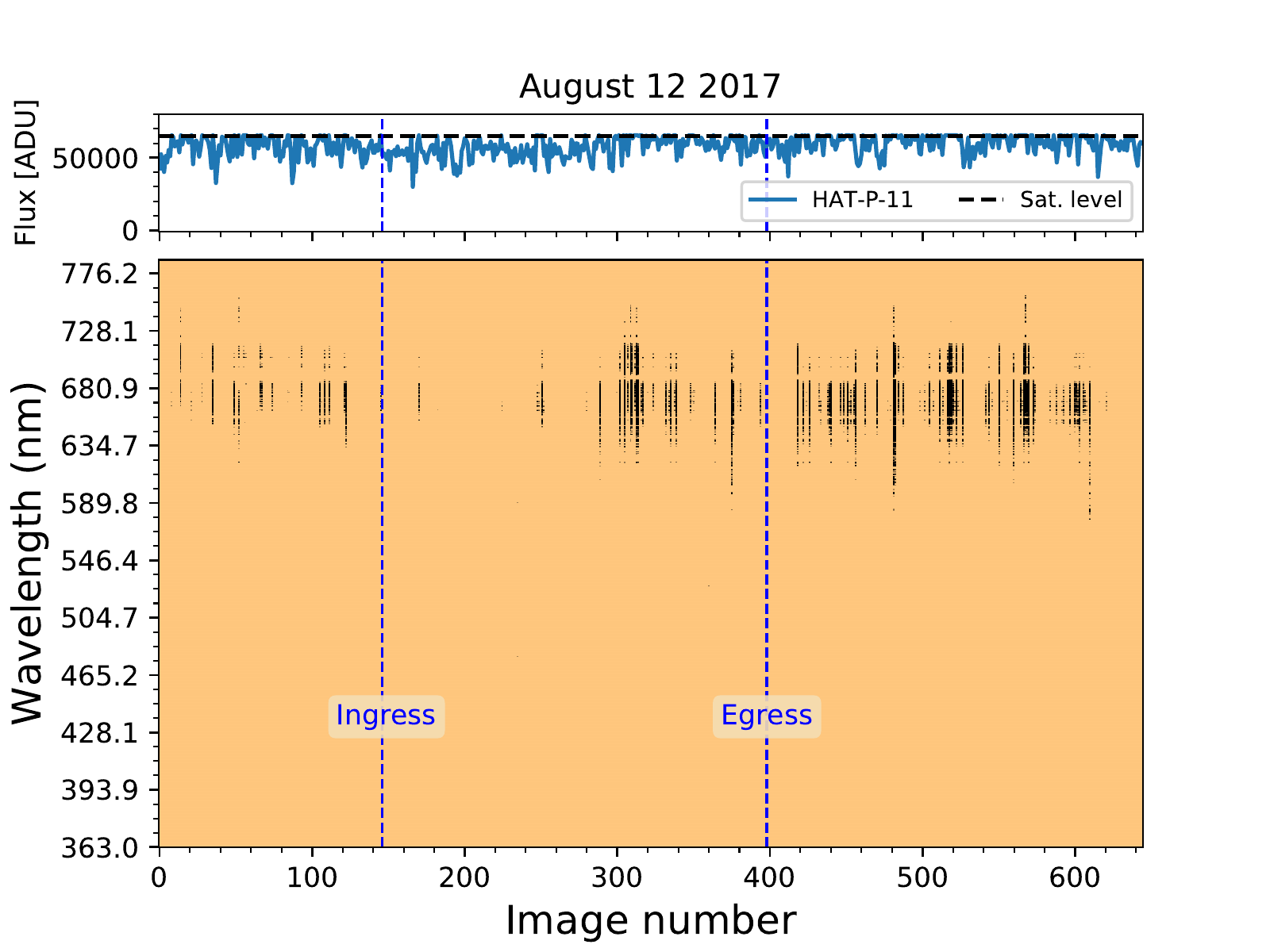}
    \caption{HAT-P-11 count level of the pixel with the maximum flux inside the extraction aperture (top) and saturated pixels of the time series (bottom) for the night of August 12 2017. In the bottom panel the saturated pixels are shown in black while the orange pixels are below the saturation level. The flux of the contaminant star was extracted using the pre-transit non-saturated images.}
    \label{Fig:Satur12Aug2017}
  \end{figure}

The night of August 12 2017 GTC observed a transit of HAT-P-11b, however the target was saturated in a large fraction of images and we decided to not use this data set to obtain the transmission spectrum of this planet. During this night the seeing conditions were good enough to separate the flux of HAT-P-11 and a nearby star, hence we used this data set to establish the flux contamination of including this nearby star inside the extraction aperture. Figure \ref{Fig:Satur12Aug2017} shows the flux level of the pixel with the maximum flux inside the extraction aperture during the night (top panel) and the saturated pixels along wavelength of the time series (bottom panel). The flux of the contaminant star was extracted using the pre-transit non-saturated images.

  \subsection{Light curve fitting tests}
  We compared the transmission spectrum of HAT-P-11b obtained by fitting the spectroscopic light curves with Gaussian Processes and without GPs for each individual nights. We tested several models with different baseline functions and parameters. For the final comparison we selected the models with the lowest Bayesian Information Criterion (BIC) value. The two non-GP baseline models selected were
  \begin{equation}
    p_1 = c_0 + c_1S_y + c_2S_y^{2} + c_3t + c_4t^{2} + c_5t^{3}
  \end{equation}
  and  
  \begin{equation}
    p_2 = c_0 + c_1S_y + c_2S_y^{2} + c_3P_y + c_4P_y^{2} + c_5t + c_6t^{2} + c_7t^{3}
  \end{equation}
  where $c_i$ are fitted coefficients, $t$ is the time, $S_y$ is the FWHM in spatial direction, and $P_y$ is the pixel position of the center of the Point Spread Function (PSF) of the spectral profile. To account for the effect of red noise in the data we used the time-averaging method (\citealp{Pont2006}).

\begin{figure}
    \centering
    \includegraphics[width=\hsize]{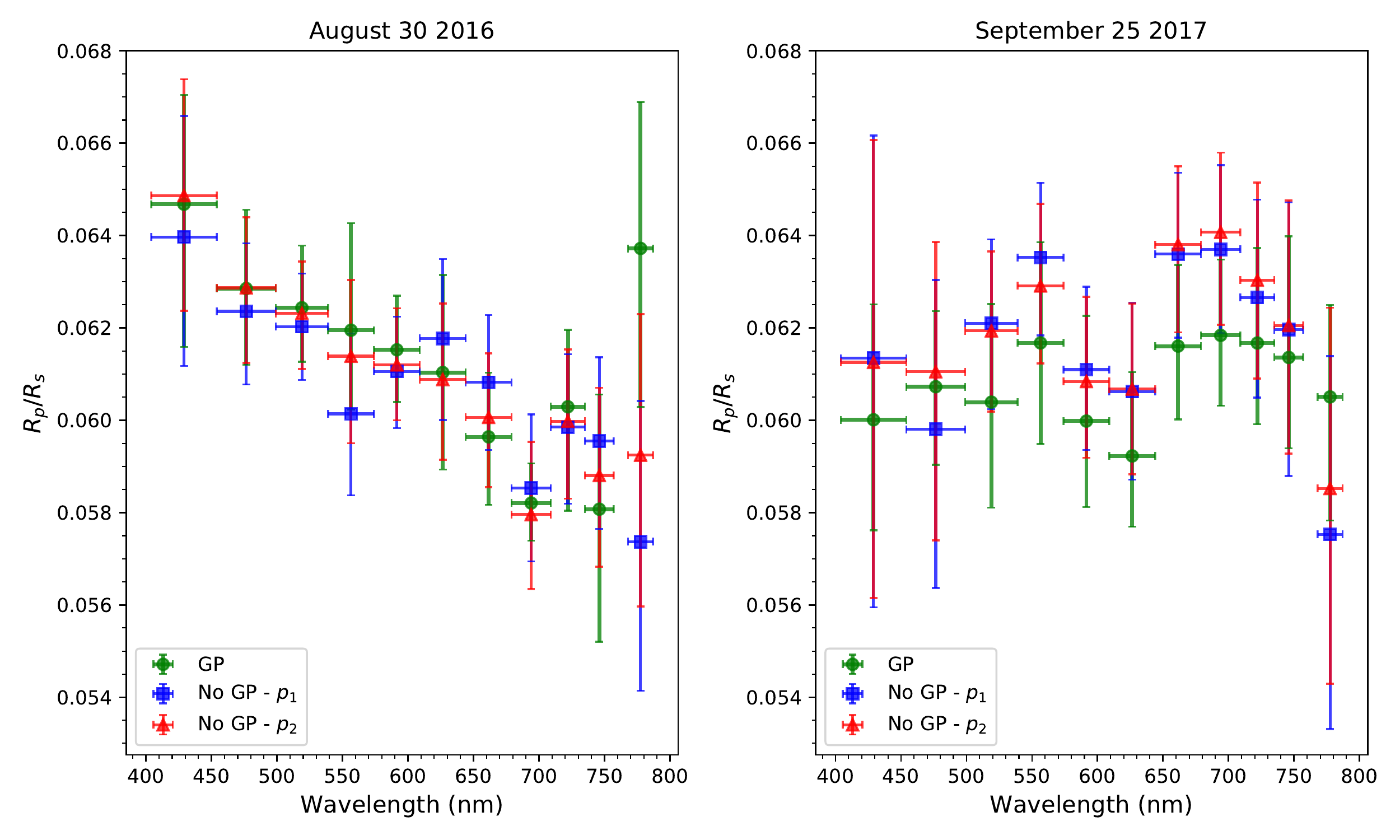}
    \caption{Comparison between fitting procedures for the individual data sets taken on August 30 2016 (left) and September 25 2017 (right). The green points are the results obtained with Gaussian Processes and the blue squares and red triangles show the results obtained without GPs and different baseline functions.}
    \label{Fig:TransSpec_GP_vs_noGP}
  \end{figure}

  Figure \ref{Fig:TransSpec_GP_vs_noGP} presents the transmission spectrum obtained with and without GPs for the baseline models $p_1$ and $p_2$. Both methods show consistent results and the slope in the transmission spectrum does not change significantly.
  
  \subsection{Spectroscopic light curves}
    In this section of the appendix we present the spectroscopic light curves of HAT-P-11b for the nights of August 30 2016 and September 25 2017.

    \begin{figure*}
      \centering
      \includegraphics[width=\hsize]{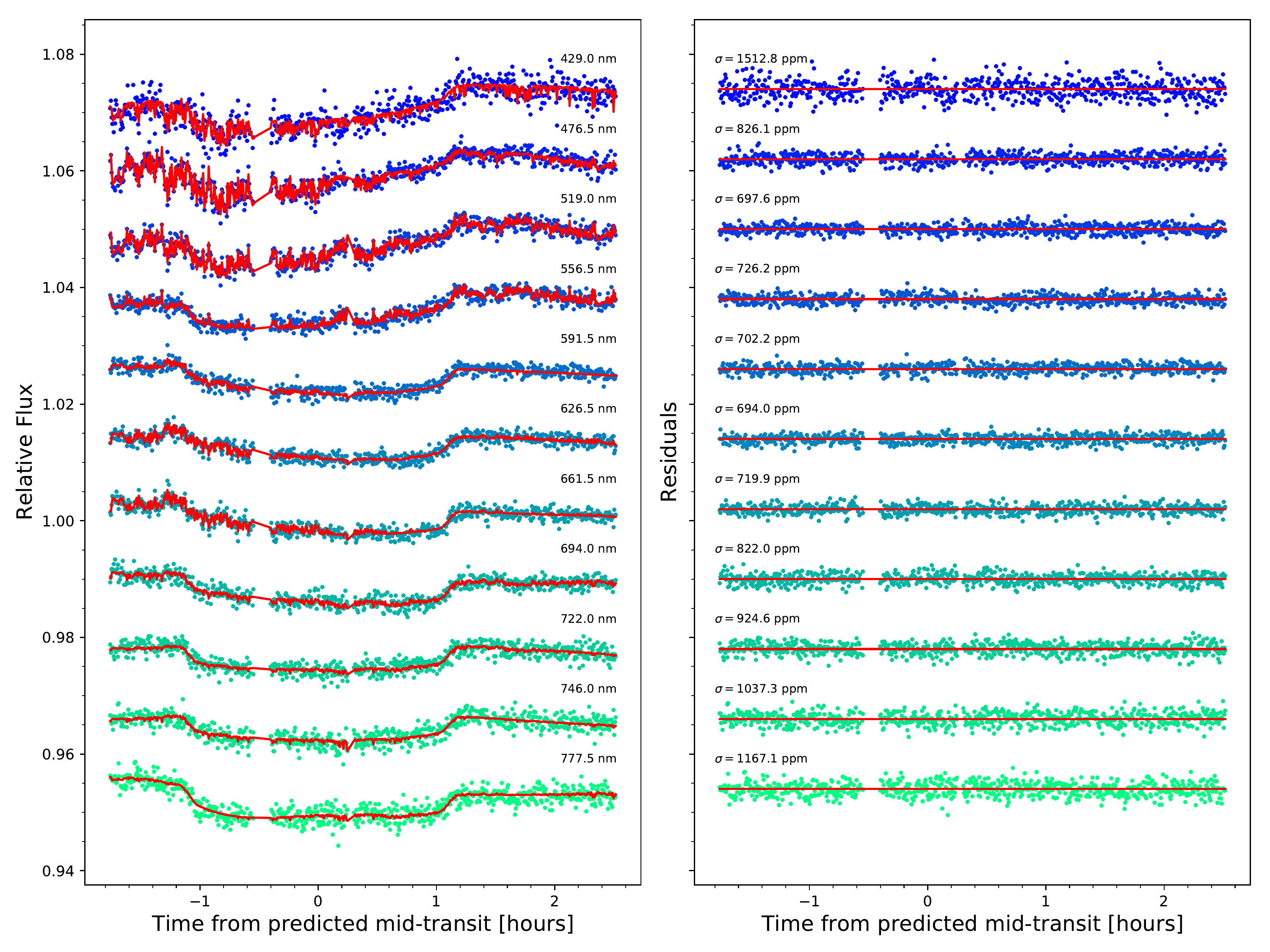}
      \caption{Spectroscopic light curves obtained using the custom passbands (see Fig. \ref{Fig:StarSpec}) for HAT-P-11b transit of August 30 2016. \textit{Left panel}: light curves after removing the common-mode systematics and best fit model (red line). \textit{Right panel}: Residuals of the light curves after removing a transit model and systematic effects.}
      \label{Fig:BinCurves_N1}
    \end{figure*}

    \begin{figure*}
      \centering
      \includegraphics[width=\hsize]{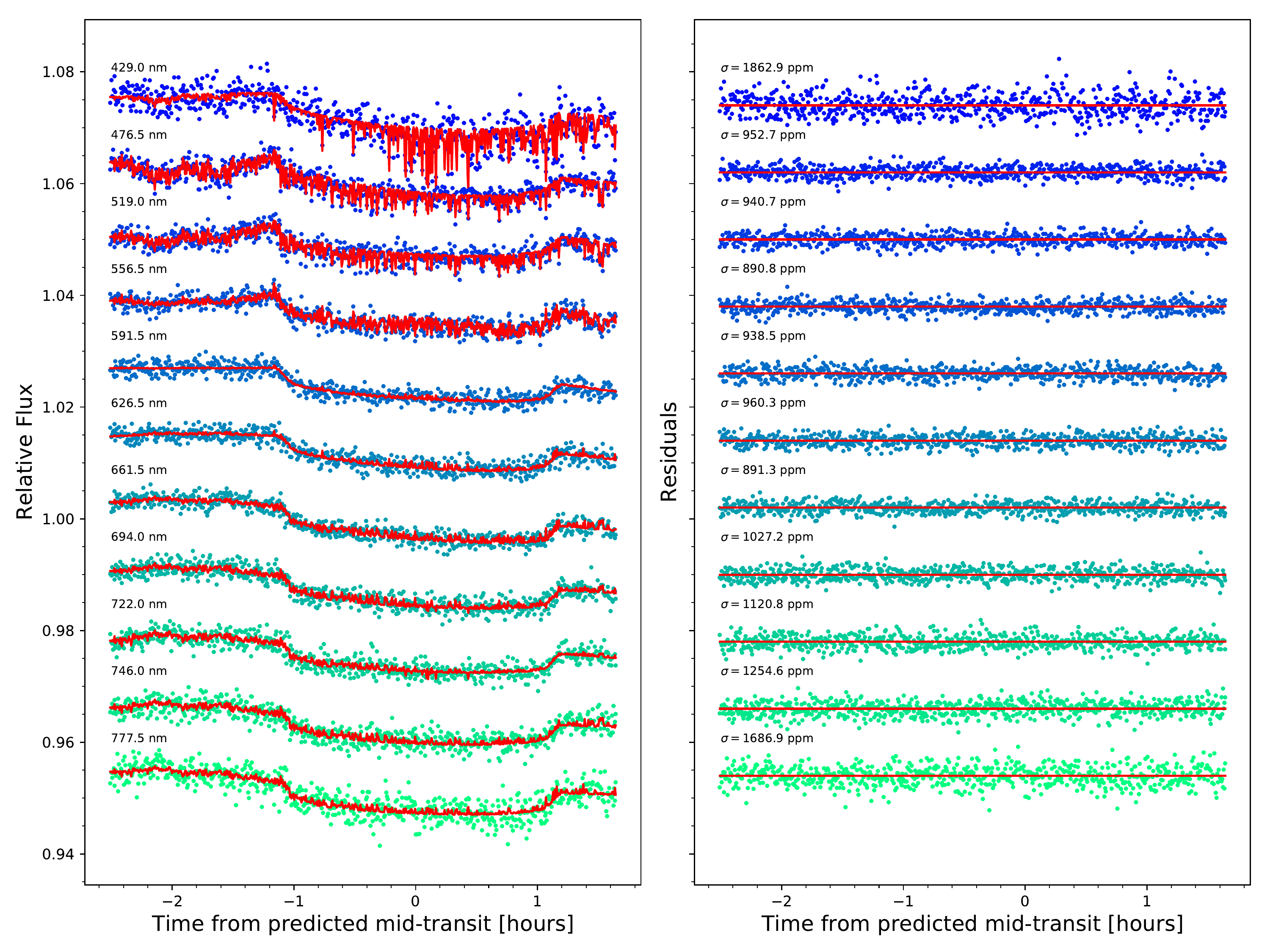}
      \caption{Spectroscopic light curves obtained using the custom passbands (see Fig. \ref{Fig:StarSpec}) for HAT-P-11b transit of September 25 2017. \textit{Left panel}: light curves after removing the common-mode systematics and best fit model (red line). \textit{Right panel}: Residuals of the light curves after removing a transit model and systematic effects.}
      \label{Fig:BinCurves_N2}
    \end{figure*}

   \subsection{Correlation plots}
   Here we present the correlation plots of the MCMC fitting procedure of Eq. \ref{Eq:RpRs_spots} for N1 and N2.
   
    \begin{figure}
      \centering
      \includegraphics[width=\hsize]{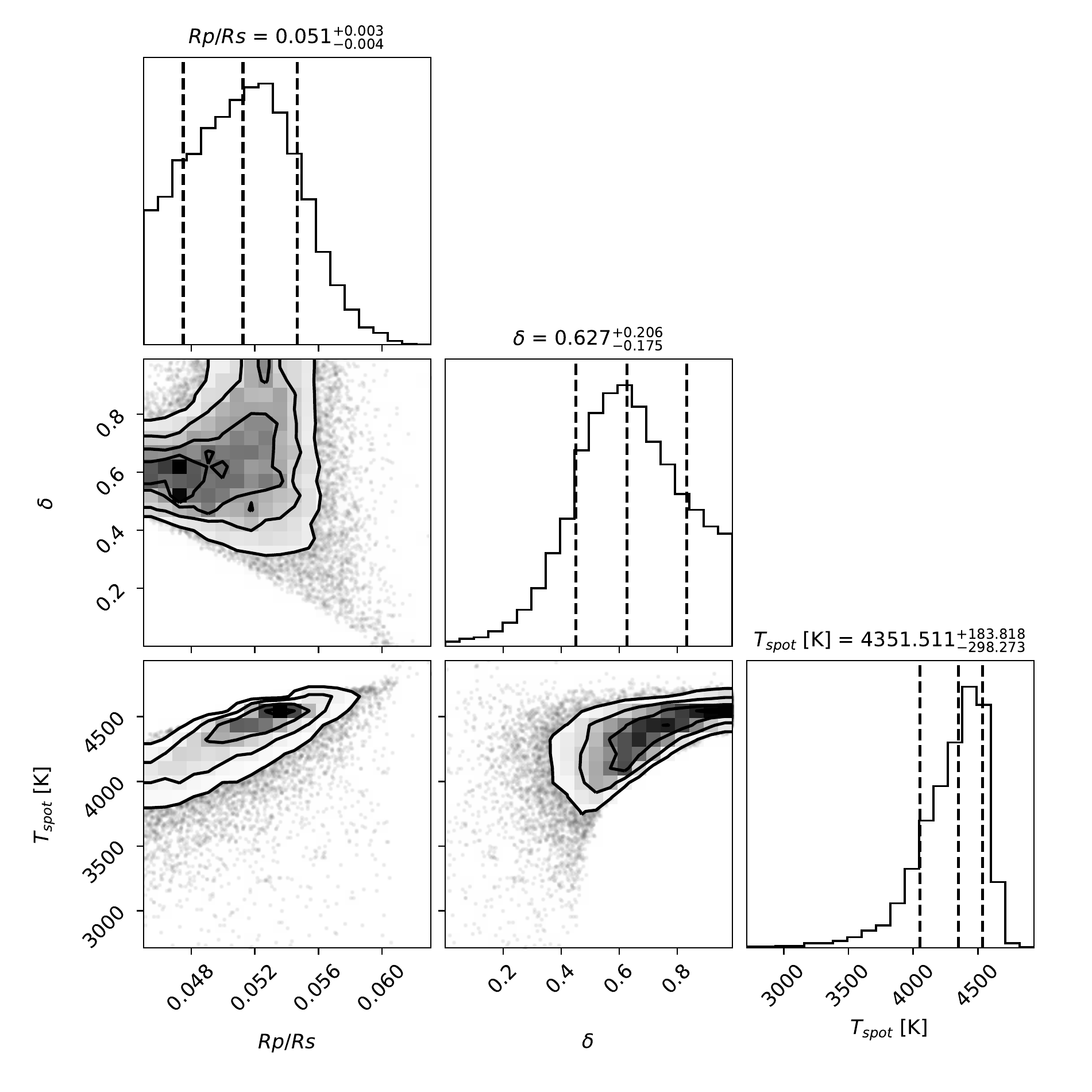}
      \caption{Correlation plot for the spot coverage model for N1.}
      \label{Fig:CorPlotN1}
    \end{figure}
    
    \begin{figure}
      \centering
      \includegraphics[width=\hsize]{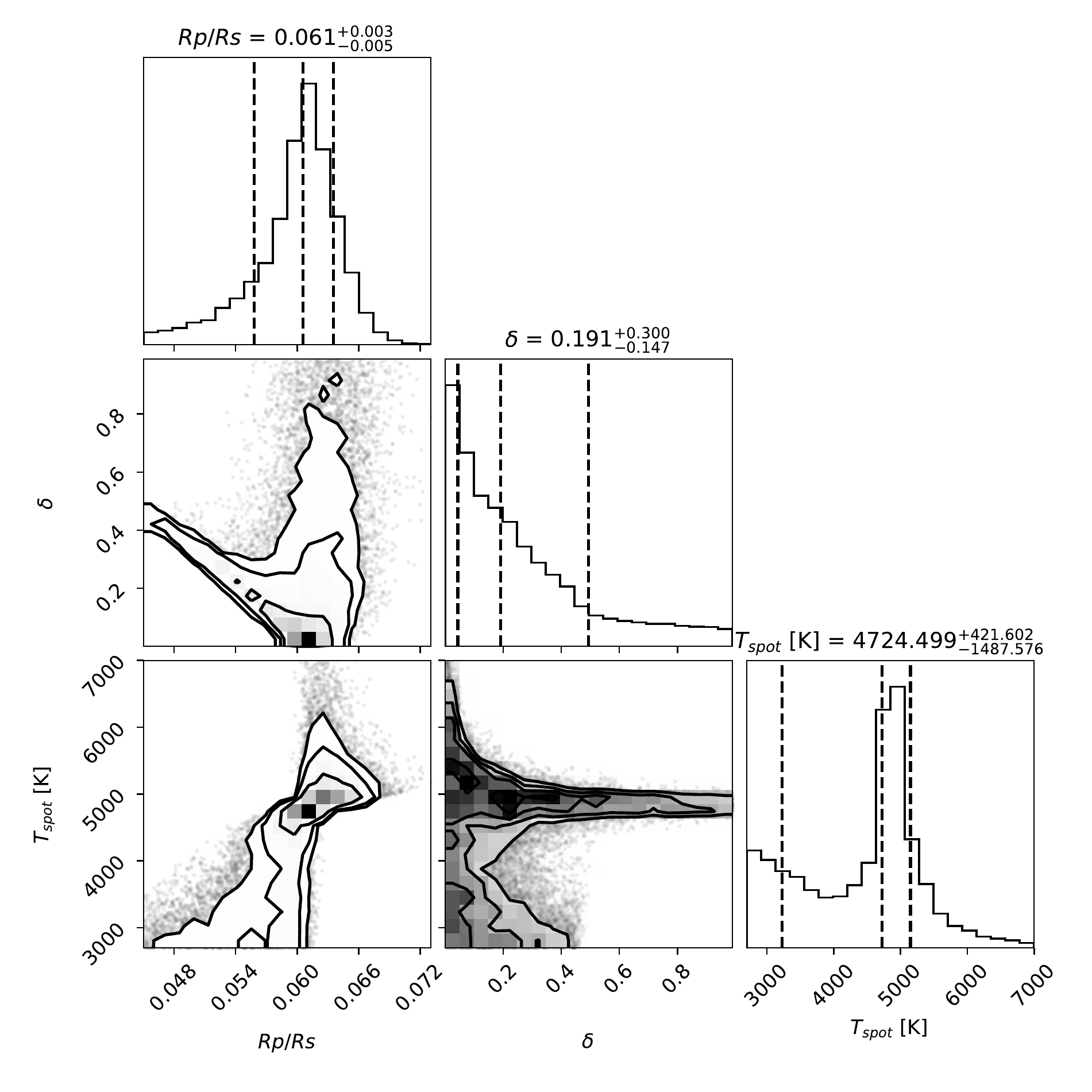}
      \caption{Correlation plot for the spot coverage model for N2.}
      \label{Fig:CorPlotN2}
    \end{figure}
    
\end{appendix}

\end{document}